\documentclass{aa}  

\usepackage{graphicx}
\usepackage{multirow}
\usepackage{txfonts}
\usepackage{xcolor}
\usepackage{comment}
\usepackage{amsmath}

\newcommand{\tsup}{\textsuperscript}
\newcommand{\tsub}{\textsubscript}
\newcommand{\LTT}{\mbox{LTT 1445A}}
\newcommand{\LTTb}{\mbox{LTT 1445Ab}}
\newcommand{\LTTc}{\mbox{LTT 1445Ac}}
\newcommand{\GJ}{\mbox{GJ 486}}
\newcommand{\GJb}{\mbox{GJ 486b}}
\newcommand{\Lya}{Ly$\alpha$}
\newcommand{\C}{CO\tsub{2}}

\usepackage{hyperref}
\hypersetup{colorlinks, 
            citecolor=blue, 
            linkcolor=blue
            }

\begin{document}

   \title{High-energy spectra of LTT 1445A and GJ 486\\reveal flares and activity}


   \author{
          H.\ Diamond-Lowe\inst{1}$^{\href{https://orcid.org/0000-0001-8274-6639}{\includegraphics[scale=0.04]{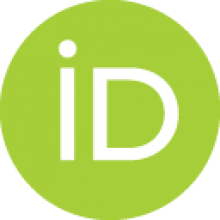}}}$
          \and
          G.\ W.\ King\inst{2}$^{\href{https://orcid.org/0000-0002-3641-6636}{\includegraphics[scale=0.04]{orcid-ID.png}}}$
          \and
          A.\ Youngblood\inst{3}$^{\href{https://orcid.org/0000-0002-1176-3391}{\includegraphics[scale=0.04]{orcid-ID.png}}}$
          \and
          A.\ Brown\inst{4}$^{\href{https://orcid.org/0000-0003-2631-3905}{\includegraphics[scale=0.04]{orcid-ID.png}}}$
          \and
          W.\ S.\ Howard\inst{5,19}$^{\href{https://orcid.org/0000-0002-0583-0949}{\includegraphics[scale=0.04]{orcid-ID.png}}}$
          \and
          J.\ G.\ Winters\inst{6,7}$^{\href{https://orcid.org/0000-0001-6031-9513}{\includegraphics[scale=0.04]{orcid-ID.png}}}$
          \and
          \mbox{D.\ J.\ Wilson}\inst{8}$^{\href{https://orcid.org/0000-0001-9667-9449}{\includegraphics[scale=0.04]{orcid-ID.png}}}$
          \and
          K.\ France\inst{4,5,8}$^{\href{https://orcid.org/0000-0002-1002-3674}{\includegraphics[scale=0.04]{orcid-ID.png}}}$
          \and
          \mbox{J.\ M.\ Mendon\c{c}a\inst{1}$^{\href{https://orcid.org/0000-0002-6907-4476}{\includegraphics[scale=0.04]{orcid-ID.png}}}$}
          \and
          L.\ A.\ Buchhave\inst{1}$^{\href{https://orcid.org/0000-0003-1605-5666}{\includegraphics[scale=0.04]{orcid-ID.png}}}$
          \and
          L.\ Corrales\inst{2}$^{\href{https://orcid.org/0000-0002-5466-3817}{\includegraphics[scale=0.04]{orcid-ID.png}}}$
          \and
          \mbox{L.\ Kreidberg}\inst{9}$^{\href{https://orcid.org/0000-0003-0514-1147}{\includegraphics[scale=0.04]{orcid-ID.png}}}$
          \and
          \mbox{A.\ A.\ Medina}\inst{10}$^{\href{https://orcid.org/0000-0001-8726-3134}{\includegraphics[scale=0.04]{orcid-ID.png}}}$
          \and
          J.\ L.\ Bean\inst{11}$^{\href{https://orcid.org/0000-0003-4733-6532}{\includegraphics[scale=0.04]{orcid-ID.png}}}$
          \and
          Z.\ K.\ Berta-Thompson\inst{5}$^{\href{https://orcid.org/0000-0002-3321-4924}{\includegraphics[scale=0.04]{orcid-ID.png}}}$
          \and
          T.\ M.\ Evans-Soma\inst{9,12}$^{\href{https://orcid.org/0000-0001-5442-1300}{\includegraphics[scale=0.04]{orcid-ID.png}}}$
          \and
          C.\ Froning\inst{13}$^{\href{https://orcid.org/0000-0001-8499-2892}{\includegraphics[scale=0.04]{orcid-ID.png}}}$
          \and
          \mbox{G.\ M.\ Duvvuri}\inst{14}$^{\href{https://orcid.org/0000-0002-7119-2543}{\includegraphics[scale=0.04]{orcid-ID.png}}}$
          \and
          \mbox{E.\ M.-R.\ Kempton}\inst{15}$^{\href{https://orcid.org/0000-0002-1337-9051}{\includegraphics[scale=0.04]{orcid-ID.png}}}$
          \and
          \mbox{Y.\ Miguel}\inst{16,17}$^{\href{https://orcid.org/0000-0002-0747-8862}{\includegraphics[scale=0.04]{orcid-ID.png}}}$
          \and
          J.\ S.\ Pineda\inst{8}$^{\href{https://orcid.org/0000-0002-4489-0135}{\includegraphics[scale=0.04]{orcid-ID.png}}}$
          \and
          C.\ Schneider\inst{18}$^{\href{https://orcid.org/0000-0002-5094-2245}{\includegraphics[scale=0.04]{orcid-ID.png}}}$
          }

   \institute{
             Department of Space Research and Space Technology, Technical University of Denmark, Elektrovej 328, 2800 Kgs.\ Lyngby, DK\\
              \email{hdiamondlowe@space.dtu.dk}
             \and
             Department of Astronomy, University of Michigan, 1085 S.\ University Ave., 323 West Hall, Ann Arbor, MI 48109, US
             \and
             Exoplanets and Stellar Astrophysics Lab, NASA Goddard Space Flight Center, Greenbelt, MD 20771, US
             \and
             Center for Astrophysics and Space Astronomy, University of Colorado, 389 UCB, Boulder, CO 80309, US
             \and
             Department of Astrophysical and Planetary Sciences, University of Colorado, 2000 Colorado Avenue, Boulder, CO 80309, US
             \and
             Center for Astrophysics $\vert$ Harvard \& Smithsonian, 60 Garden Street, Cambridge, MA 02138, US
             \and
             Bridgewater State University, 131 Summer St., Bridgewater, MA 02325, US
             \and
             Laboratory for Atmospheric and Space Physics, 1234 Innovation Drive, Boulder, CO 80303, US
             \and
             Max Planck Institute for Astronomy, K\"onigstuhl 17, 69117 Heidelberg, DE
             \and
             Department of Astronomy, The University of Texas at Austin, Austin, TX 78712, US
             \and
             Department of Astronomy \& Astrophysics, University of Chicago, Chicago, IL 60637, US
             \and
             School of Information and Physical Sciences, University of Newcastle, Callaghan, NSW, AU
             \and
             Southwest Research Institute, 6220 Culebra Rd., San Antonio, TX 78238, US
             \and
             Department of Physics and Astronomy, Vanderbilt University, Nashville, TN 37235, US
             \and
             Department of Astronomy, University of Maryland, 4296 Stadium Drive, College Park, MD 20742, US
             \and
             Leiden Observatory, Leiden University, P.O. Box 9513, 2300 RA Leiden, NL
             \and
             SRON Netherlands Institute for Space Research, Niels Bohrweg 4, 2333 CA Leiden, NL
             \and
             Hamburger Sternwarte, Gojenbergsweg 112, D-21039, Hamburg, DE
             \and
             NHFP Sagan Fellow
             }

   \date{Received}

 
  \abstract
  {The high-energy radiative output, from the X-ray to the ultraviolet, of exoplanet host stars drives photochemical reactions and mass loss in the upper regions of planetary atmospheres. In order to place constraints on the atmospheric properties of the three closest terrestrial exoplanets transiting M dwarfs, we observe the high-energy spectra of the host stars \LTT\ and \GJ\ in the X-ray with \textit{XMM-Newton} and \textit{Chandra} and in the ultraviolet with HST/COS and STIS. We combine these observations with estimates of extreme-ultraviolet flux, reconstructions of the \Lya\ lines, and stellar models at optical and infrared wavelengths to produce panchromatic spectra from 1 \AA\ to 20 $\mu$m for each star. While \LTTb, \LTTc, and \GJb\ do not possess primordial hydrogen-dominated atmospheres, we calculate that they are able to retain pure \C\ atmospheres if starting with 10, 15, and 50\% of Earth's total \C\ budget, respectively, in the presence of their host stars' stellar wind. We use age--activity relationships to place lower limits of 2.2 and 6.6 Gyr on the ages of the host stars \LTT\ and \GJ. Despite both \LTT\ and \GJ\ appearing inactive at optical wavelengths, we detect flares at ultraviolet and X-ray wavelengths for both stars. In particular, \GJ\ exhibits two far-ultraviolet flares with absolute energies of 10\tsup{29.5} and 10\tsup{30.1} erg (equivalent durations of $4357\pm96$ and $19724\pm169$ s) occurring three hours apart. Based on the timing of the observations, we suggest that these high-energy flares are related and indicative of heightened flaring activity that lasts for a period of days, but our interpretations are limited by sparse time-sampling. Consistent high-energy monitoring is needed to determine the duration and extent of high-energy activity on individual M dwarfs and the population as a whole.}


  \titlerunning{Spectra of LTT 1445A and GJ 486}
  \authorrunning{Diamond-Lowe, et al.}

  \maketitle

\section{Introduction}

The Kepler mission \citep{Borucki2010}, which operated from 2009 to 2013, determined that small planets ($<4\ \mathrm{R_\oplus}$) are the most abundant in the Galaxy \citep{Fressin2013,Fulton2017}. The Transiting Exoplanet Survey Satellite \citep[TESS, launched 2018 and continuing to operate;][]{Ricker2015} along with ground-based transit and radial velocity facilities detect small planets orbiting our closest stellar neighbors and determine the radii and masses of these worlds \citep[e.g.,][]{Mayor2003,Nutzman2008,Quirrenbach2010,Mahadevan2010,Cosentino2012,Gillon2013,Irwin2015,Seifahrt2016,Schwab2016,Bouchy2017,Artigau2014}. Now, the \textit{James Webb Space Telescope} (JWST, launched 2021) is providing unprecedented insight into the atmospheres of small worlds \citep[e.g.,][]{Lustig-Yaeger2023,Greene2023,Zieba2023,Moran2023,Lim2023,May2023,Zhang2024}. 

A common thread among all small planets whose atmospheres we are currently able to study with state-of-the-art observatories is that they orbit M dwarfs, a stellar type with masses and radii below 0.6$\times$ that of the Sun and effective temperatures below 3900 K. While the small sizes, low masses, and large population of nearby M dwarfs allow for relative ease of small planet detection, their extended pre-main sequence phases, persistent activity and flaring, and high-energy fluxes threaten to permanently alter or destroy the atmospheres of those small planets \citep[e.g.,][]{Segura2010,Seager2010,Teal2022,Chen2021,Howard2023}. 

Stellar ultraviolet photons drive photochemistry in the upper atmospheres of planets, where remote sensing techniques of exoplanet atmospheres, such as transmission spectroscopy, are most sensitive. The first evidence for photochemistry outside the Solar System was recently detected in the atmosphere of a hot Jupiter \citep{Tsai2023}, demonstrating the importance of understanding complex photochemical reactions when interpreting planetary atmospheres. The relative proportions of far- to near-UV (FUV=912--1700 \AA; NUV=1700--3200 \AA) flux are particularly important for determining the balance of molecular species, such as H\tsub{2}O, CH\tsub{4}, CO\tsub{2}, CO, O\tsub{2}, and O\tsub{3}, in planetary atmospheres \citep{Tian2014,Harman2015,Rugheimer2015}.

High-energy flux at X-ray (=1--100 \AA) and extreme-ultraviolet (EUV=100--912 \AA) wavelengths drives atmospheric mass loss and has the potential to completely strip small planets of their primordial atmospheres \citep{Lopez2012,Owen&Wu2017}. This process occurs during the highly active saturation phase of young stars on timescales of around 100 Myr \citep{Lopez2013,Owen&Wu2013}, but mass loss from EUV photons may persist on gigayear timescales \citep{King&Wheatley2021}, and complete atmospheric stripping can continue out to 10 Gyr with active flaring \citep{France2020}. Small planets on tight orbits around M dwarfs are also vulnerable to atmospheric stripping by stellar winds \citep{Cohen2015,Garraffo2016,Garraffo2017}.

Results from the large MUSCLES program demonstrated that M dwarfs that are similar in size, mass, and temperature produce a range of fluxes in the UV, making the scaling of high-energy flux from one M dwarf to another highly uncertain \citep{France2016,Youngblood2017,Melbourne2020}. When determining whether or not a terrestrial planet retains an atmosphere, having the exact high-energy output from its host M dwarf is essential, as is work to determine its high-energy past. In the present work, we provide a snapshot of the UV and X-ray outputs of the two closest M dwarfs known to host transiting terrestrial planets: \LTT\ and \GJ. To capture the high-energy spectra of these stars, we use the unique spectral coverage and resolving power of the \textit{Hubble} Space Telescope coupled with X-ray information from \textit{XMM-Newton} and \textit{Chandra}. Both \LTT\ and \GJ\ are considered to be old, inactive M dwarfs based on their measured rotation periods and optical activity indicators, such as H$\alpha$ and Ca \textsc{ii} H \&\ K. However, M dwarfs that are quiet at optical wavelengths are known to flare at higher energies \citep[e.g.,][]{Loyd2018,Jackman2024}. A flare from \LTT\ at X-ray wavelengths has already been reported in \citet{Brown2022}. 

At the time of writing, there are two known planetary companions orbiting \LTT\ and one planetary companion orbiting \GJ\ \citep{Winters2019,Trifonov2021,Winters2022,Caballero2022}. All three planets have masses and radii consistent with terrestrial bulk compositions (Table~\ref{tab:both_stellar_params}). Ground-based optical transmission spectroscopy of \LTTb\ rules out solar-composition atmospheres down to 1 bar of surface pressure \citep{Diamond-Lowe2023}. Ground-based high-resolution transmission spectroscopy of \GJb\ rules out clear H/He-dominated atmospheres as well as clear 100\% water vapor atmospheres \citep{Ridden-Harper2023}. JWST transmission spectroscopy of \GJb\ rules out much higher mean-molecular-weight atmospheres at 1 bar of surface pressure \citep{Moran2023}. The same JWST data may suggest a water-rich atmosphere on this planet; however the interpretation of these results is degenerate with the presence of star-spots in the transmission spectrum. Upcoming results from JWST emission spectroscopy (GO 1743, PI Mansfield; and 2807 PI Berta-Thompson) and transmission spectroscopy (GO 2515, PI Batalha) will provide further clues as to the atmospheric status of \LTTb\ and \GJb. To complete the atmospheric picture of these planets, we need measurements of the high-energy spectra of their host stars to pass into photochemical and mass-loss models. Whether or not these planets have atmospheres, and what those atmospheres or rocky surfaces look like, depends on their high-energy stellar environments \citep{Louca2023}. 

The present paper is laid out as follows: In Sect.\,\ref{sec:obs} we provide information on the observations that go into this work. In Sect.\,\ref{sec:timeseries} we present the time-series analysis of our data. In Sect.\,\ref{sec:spectra} we put all of the measured and estimated fluxes together to build a panchromatic spectrum of our targets. We provide an analysis of detected flares in Sect.\,\ref{sec:flare}. We discuss our results pertaining to the planets orbiting \LTT\ and \GJ, as well as a discussion of M dwarf ages and activity in Sect.\,\ref{sec:discussion}. We conclude with Sect.\,\ref{sec:conclusions}. The resulting spectra of \LTT\ and \GJ\ are available for download as high-level science products (HLSPs) from the \texttt{mstarpanspec} page\footnote{doi: \href{https://dx.doi.org/10.17909/t9-fqky-7k61}{10.17909/t9-fqky-7k61}}.

\section{Observations}\label{sec:obs}

We observed the ultraviolet to optical spectra, from 1065--5700 \AA, of the two closest M dwarfs to host transiting terrestrial exoplanets, \LTT\ and \GJ. We used a combination of HST/COS and HST/STIS (GO 16722, PI H.\,Diamond-Lowe and Co-PI G.\,King; GO 16701, PI A.\,Youngblood and Co-PI K.\,France) to make these observations and achieved almost complete spectral coverage from the FUV to the optical. 

To capture the FUV (912--1700 \AA) we used COS/G130M with a central wavelength of 1222 \AA\ and COS/G160M with a central wavelength of 1533 \AA. Using COS/G130M at 1222 \AA\ places the \Lya\ line in the gap between the A and B segments, which is required as part of the bright object protections for COS. We captured NUV (1700--3200 \AA) flux using COS/G230L with a central wavelength of 2950 \AA. While COS provides higher sensitivity to key transition region lines in the FUV and NUV, the segment gap in the COS/G230L NUV spectrum is rather large, spanning 2113--2785 \AA. To fill in this gap we used STIS/G230L with a central wavelength of 2376 \AA. In order to connect the UV spectra to optical spectra, we took a brief observation with STIS/G430L with a central wavelength of 4300 \AA.

Though \Lya\ emission makes up about 85\% of FUV flux for M dwarfs \citep{France2016}, the bulk of the \Lya\ line core is absorbed by neutral hydrogen in the interstellar medium before we can observe it. For nearby bright M dwarfs it is possible to reconstruct \Lya\ flux from the observed red and blue wings of the \Lya\ line \citep{Youngblood2016}, but for most inactive M dwarfs it is prohibitively expensive to capture enough signal in the the \Lya\ line wings to perform a reconstruction. Though \LTT\ and \GJ\ are considered inactive M dwarfs, they are close enough to capture enough \Lya\ flux in the wings of the \Lya\ profile to perform the reconstruction. To make these observations we used STIS/G140M with a central wavelength of 1222 \AA\ and a narrow 52$\times$0.2$''$ slit to avoid geocoronal \Lya\ contamination, which is prevalent at \Lya\ wavelengths and would leak into the larger $2.5''$ aperture of COS. For both targets we gathered STIS/G140M data over three orbits, combining observations from the GO 16701 and 16722 programs. These observations were inspected separately but combined from the two GO Programs in order to boost S/N for the \Lya\ reconstruction. 

For the highest-energy part of the spectrum, we used X-ray data from \textit{XMM-Newton} and the \textit{Chandra} X-ray Observatory. \textit{Chandra} has the high spatial resolution necessary to separate \LTT\ from the more active LTT 1445BC binary companion 7$''$ away. With a PSF of 15$''$ \textit{XMM-Newton} cannot resolve \LTT\ from its companion stars. For \LTT\ we used X-ray measurements from the \textit{Chandra} ACIS-S instrument described in \citet{Brown2022}, as well as an additional set of observations taken in 2023 (Obs IDs 23377, PI Brown; 27882, PI Howard). 

For \GJ\ we used \textit{XMM-Newton} since it can provide broader spectral coverage in the X-ray. We observed \GJ\ for 30 ks with \textit{XMM-Newton} and the EPIC-pn camera as part of GO 16722 (PI Diamond-Lowe and Co-PI King). The apparent visual magnitude of \GJ\ is faint enough that we used all three EPIC cameras with the thin optical blocking filters and in full frame mode. We also made use of the simultaneous Optical Monitor (OM) observation taken with the UVW1 filter; however there were issues with these data, as outlined in Sect.\,\ref{ssec:xrayFlare}. We also obtained \textit{Chandra} HRC-I data for \GJ\ (31.5 ks; Obs IDs 26210, 27799, 27942, PI Youngblood) which provided additional measurements of the quiescent soft-X-ray emission. All these instruments sample similar energy ranges between 0.1 and 10 keV (corresponding to a wavelength range of 1.2--120 \AA), with lower energy limits of 0.1, 0.16, and 0.3 keV for \textit{Chandra} HRC-I, \textit{XMM-Newton}, and \textit{Chandra} ACIS respectively. However, the \textit{Chandra} ACIS detector has lost significant sensitivity below 1 keV because of molecular contamination.

No currently operating observatory can capture EUV (100--912 \AA) data of our targets; however this wavelength range is critical for estimates of atmospheric escape rates \citep{King&Wheatley2021}. We therefore used a differential emission measure \citep[DEM;][]{Duvvuri2021} to estimate the EUV flux from \LTT\ and \GJ\ from measured flux at UV and X-ray wavelengths. 

\section{Time series of \LTT\ and \GJ}\label{sec:timeseries}
\begin{figure*}
\centering
\includegraphics[width=17cm]{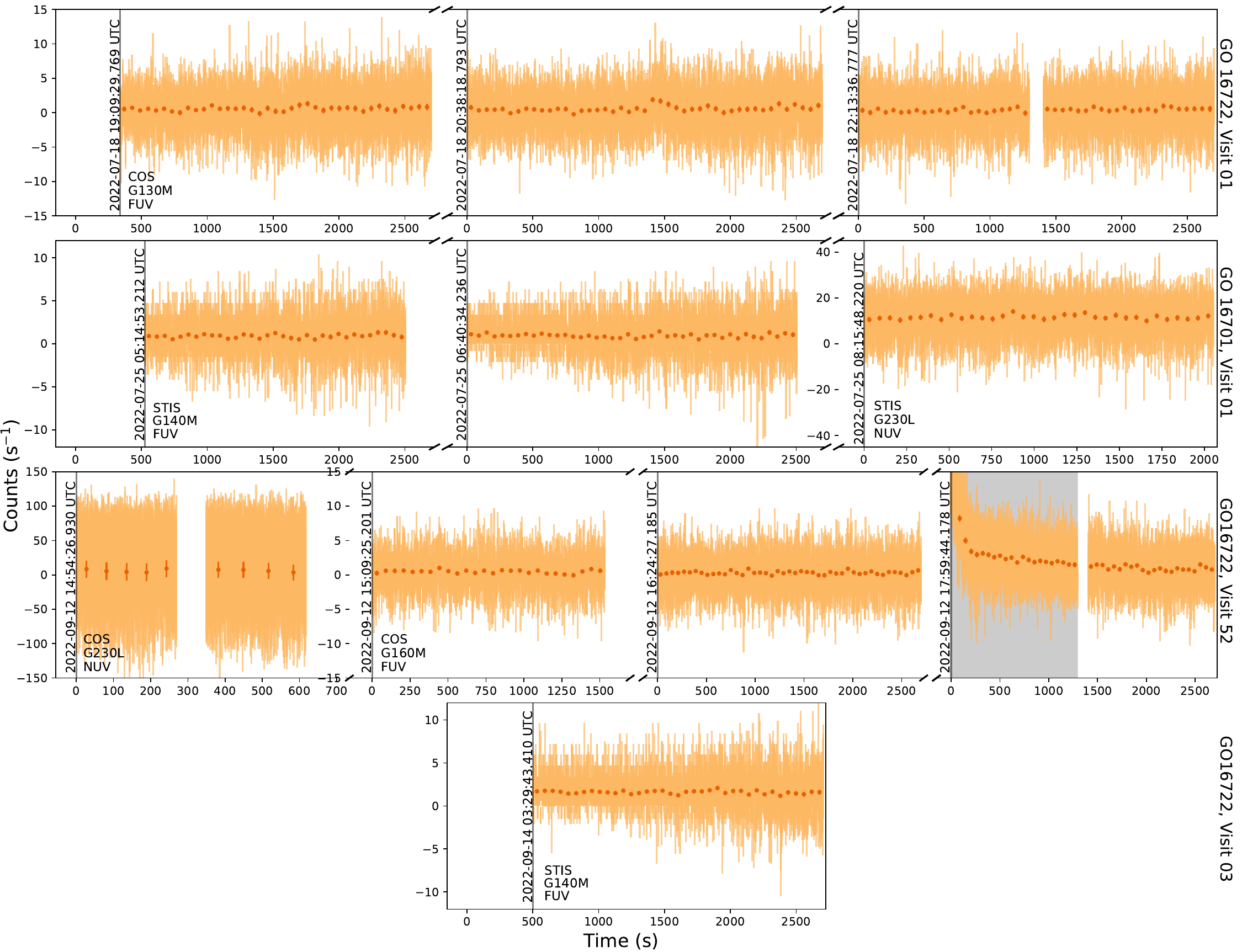}
\caption{Time series of \LTT\ observations used in this analysis from GO Programs 16722 and 16701 (program and visit number specified on right-hand side) presented in chronological order. Count rates are provided in 1 s time bins (light orange) and 1 min time bins (dark orange). The instrument, grating, and detector are given in the first panel of a set of exposures. The shaded gray region shows a flare; these data are excluded from the panchromatic spectrum (Sect.\,\ref{sec:spectra}) but  we analyze the flare data separately in Sect.\,\ref{sec:flare}.}
\label{fig:timeseries_LTT1445A}
\end{figure*}

\begin{figure*}
\centering
\includegraphics[width=17cm]{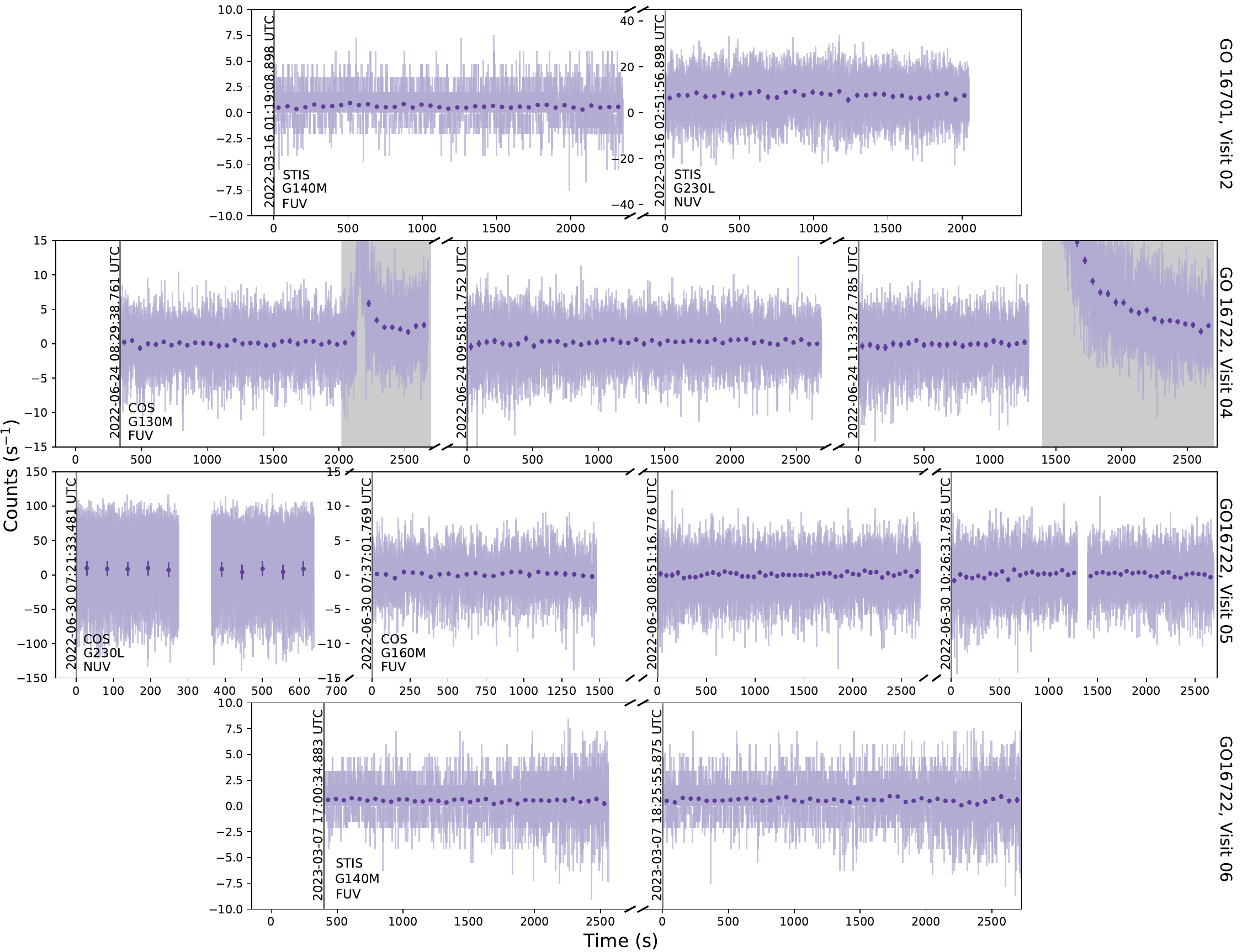}
\caption{Same as Fig.\,\ref{fig:timeseries_LTT1445A} but for \GJ.}
\label{fig:timeseries_GJ486}
\end{figure*}

All HST data used in this work, with the exception of the STIS/G430L observations, were taken in \texttt{TIME-TAG} mode, meaning that we can turn these observations into time series. We used the \texttt{*corrtag*.fits} in the case of COS and \texttt{*tag*.fits} files in the case of STIS, which have the time and detector location of each detected photon. We present the resulting time series of \LTT\ and \GJ\ in Figs.\,\ref{fig:timeseries_LTT1445A} and\,\ref{fig:timeseries_GJ486}, respectively. Negative counts are due to imperfect background subtraction. 

We detected by eye one flare in the \LTT\ COS/G160M data set and two flares in the \GJ\ COS/G130M data set (shaded regions in Figs.\,\ref{fig:timeseries_LTT1445A} and~\ref{fig:timeseries_GJ486}). As in \citet{Diamond-Lowe2021}, we used the \texttt{costools} \texttt{splittag} and \texttt{x1dcorr} functions to create our own \texttt{x1d} files that separate out the flare data from the quiescent data. We combined all \texttt{x1d} quiescent data together to create our panchromatic spectra (Sect.\,\ref{sec:spectra}). Flare data from multiple flares were not combined; each flare was processed separately to determine flare properties (Sect.\,\ref{sec:flare}). We additionally detected what may be a small flare in the \LTT\ COS/G130M observations (top middle panel of Fig.\,\ref{fig:timeseries_LTT1445A}); however our flare analysis (discussed in depth in Sect.\,\ref{sec:flare}) did not find the flux to be statistically different from the baseline flux. We did not exclude this data from the spectral analysis. 

\section{Panchromatic spectra of \LTT\ and \GJ}\label{sec:spectra}

To produce panchromatic spectra of \LTT\ and \GJ\ we combined HST UV measurements, X-ray measurements from \textit{XMM-Newton} and \textit{Chandra X-ray Observatory}, an estimation of the EUV output using the differential emission measure (DEM) method, a reconstruction of the \Lya\ line, and an extension of the spectra into the infrared with BT-Settl (CIFIST) models \citep{Allard2003,Caffau2011}. Following previous works \citep{France2016,Loyd2016,Diamond-Lowe2021,Diamond-Lowe2022}, we outline the steps to producing our panchromatic spectra here. High level science products are available on the \texttt{mstarpanspec} page of the MAST archive\footnote{\href{https://archive.stsci.edu/hlsp/mstarpanspec}{https://archive.stsci.edu/hlsp/mstarpanspec}}. In this section we work only with the quiescent data, after having removed the flare data during the time series analysis (Sect.\,\ref{sec:timeseries}). We provide stellar parameters used throughout this work in Table~\ref{tab:both_stellar_params}. 
\begin{table}
\centering
\caption{Parameters for the \LTT\ and \GJ\ systems used in this work \label{tab:both_stellar_params}}
\setlength{\tabcolsep}{1pt}
\begin{tabular}{lc|cc|c}
\hline\hline
Star         &  Unit & \multicolumn{2}{c|}{LTT 1445A}    & GJ 486   \\ 
\hline
Distance       &  pc   & \multicolumn{2}{c|}{$6.8638\pm0.0012$} & $8.0791\pm0.0021$    \\
Radius & $R_\odot$ & \multicolumn{2}{c|}{$0.265^{+0.011}_{-0.010}$}  & $0.328\pm0.011$    \\
Mass   & $M_\odot$ & \multicolumn{2}{c|}{$0.257\pm0.014$} & $0.323\pm0.015$     \\
$T_\text{eff}$ & K & \multicolumn{2}{c|}{$3340\pm150$}    & $3340\pm54$     \\
\hline
Planet           &            & b & c &           b  \\
\hline
Radius    & $R_\oplus$ & $1.305^{+0.066}_{-0.061}$   & $1.147^{+0.055}_{-0.054}$   & $1.305^{+0.063}_{-0.067}$   \\
Mass      & $M_\oplus$ & $2.87^{+0.26}_{-0.25}$   & $1.54^{+0.20}_{-0.19}$  & $2.82^{+0.11}_{-0.12}$   \\
Density   & g cm$^{-3}$ & $7.1^{+1.2}_{-1.1}$        & $5.57^{+0.68}_{-0.60}$    & $7.0^{+1.2}_{-1.0}$ \\
$a$       & AU         & $0.03813^{+0.00068}_{-0.00070}$  & $0.02661^{+0.00047}_{-0.00049}$  & $0.01734^{+0.00026}_{-0.00027}$ \\
$T_\text{eq}$ & K      & $424\pm21$    & $508\pm25$    & $701\pm13$  
\end{tabular}
\tablefoot{Distance values are from \textit{Gaia} Data Release 3 \citep{GaiaDR32023}, otherwise values for the \LTT\ system are from \citet{Winters2022} and values for \GJ\ are from \citet{Trifonov2021}.
 }
\end{table}

\subsection{Ultraviolet}\label{subsec:ultraviolet}

We directly measured the far- and near-UV output of \LTT\ and \GJ\ with HST/COS and STIS. Between these two instruments we have almost complete coverage of the UV spectrum, with the exception of the \Lya\ line and regions contaminated by geocoronal airglow. 

Using a list of preidentified geocoronal emission lines\footnote{\href{https://www.stsci.edu/hst/instrumentation/cos/calibration/airglow}{https://www.stsci.edu/hst/instrumentation/cos/calibration/airglow}}, we identified contamination in our spectra from N \textsc{i} at 1134.980\,\AA\ and 1200\,\AA\ and from O \textsc{i} at 1302--1307\,\AA\ and 1355.6\,\AA\ \citep{Feldman2001}. We zeroed-out flux in regions where we identify airglow contamination in the stellar spectra. We note that there are models and tools for modeling and removing contamination from O \textsc{i} airglow lines at 1302--1307 \AA\ \citep{Bourrier2018,CruzAguirre2023}; however we found that modeling and removing the O \textsc{i} contamination can introduce spurious flux into our spectra, while we saw no evidence of underlying stellar spectral features that needed to be preserved.

\begin{figure*}
\centering
\includegraphics[width=17cm]{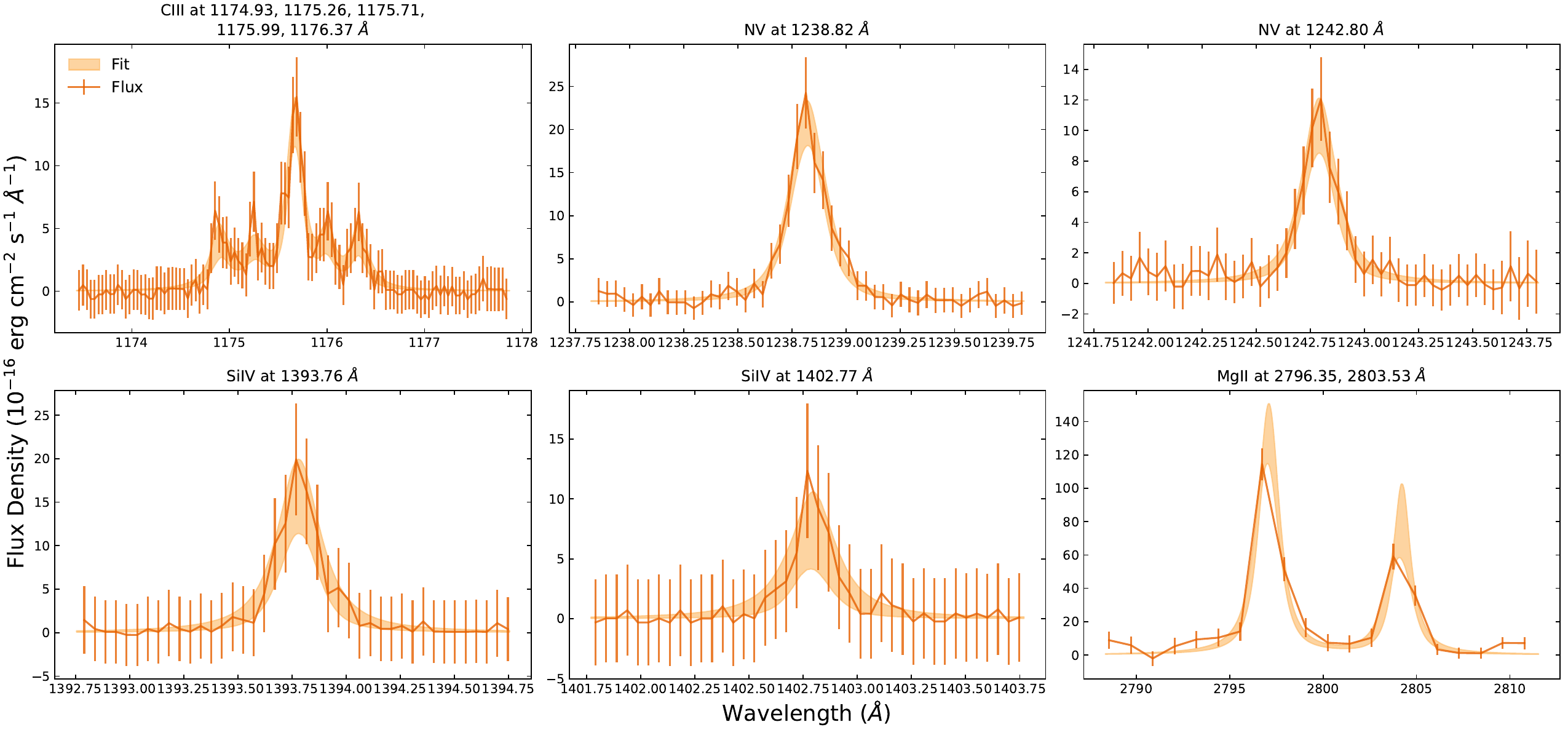}
\caption{Sample of measured UV lines with fitted Voigt models for \LTT. The sample covers UV lines from all three COS gratings used (G130M, G160M, and G230L). Lines that are blended, e.g., C \textsc{iii} and Mg \textsc{ii,} were fit with combined Voigt models, where as well-separated lines, e.g., N \textsc{v} and Si \textsc{iv,} were fit with individual models. Voigt models are made with \texttt{pyspeckit} \citep{Ginsburg2022} and we explore the model parameter space with \texttt{dynesty} \citep{Speagle2020}. }
          \label{fig:lines_LTT1445A}%
\end{figure*}

\begin{figure*}
\centering
\includegraphics[width=17cm]{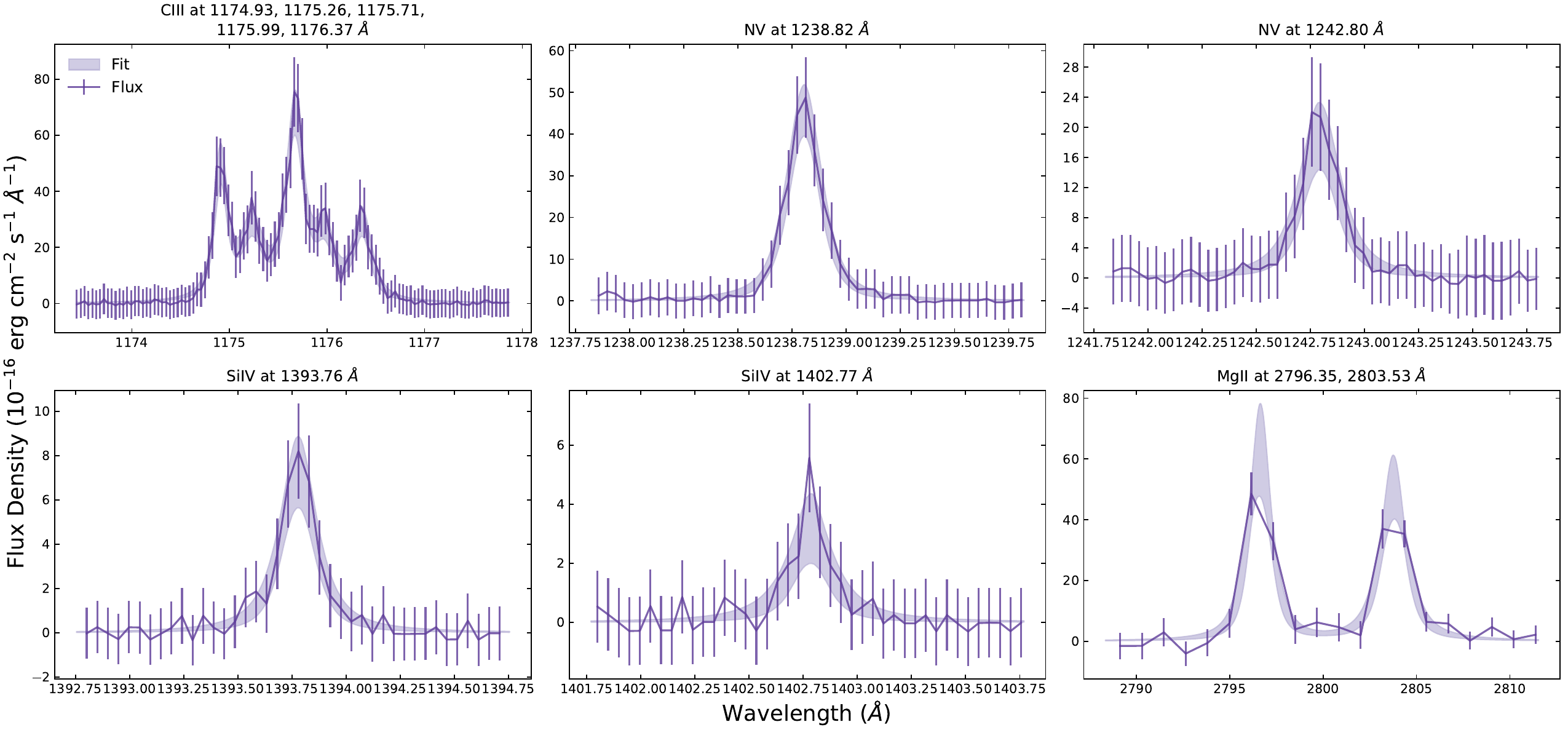}
\caption{Same as Fig.\,\ref{fig:lines_LTT1445A} but for \GJ.} 
          \label{fig:lines_GJ486}%
\end{figure*}

We measured the flux in prominent UV transition region lines by fitting Voigt models convolved with the COS line spread function (LSF) corresponding to the correct lifetime position for the data to each spectral line. To redefine the LSF from pixels to wavelength we followed instructions in the STScI COS Jupyter notebook on working with COS LSFs\footnote{\href{https://spacetelescope.github.io/COS-Notebooks/LSF.html}{https://spacetelescope.github.io/COS-Notebooks/LSF.html}}. We used the open-source \texttt{pyspeckit} \citep{Ginsburg&Mirocha2011,Ginsburg2022} to construct the Voigt model, and we estimated uncertainties in the model fit by exploring the parameter space with the open-source \texttt{dynesty} nested sampler \citep{Speagle2020}. For blended lines, such as C \textsc{iii}, we fit a combination of Voigt models simultaneously. For lines with well-separated peaks, such as N \textsc{v}, we fit each peak separately, and then summed them for further analysis. We provide a sample of fitted line fluxes in Figs.\,\ref{fig:lines_LTT1445A} and~\ref{fig:lines_GJ486}. 

Integrated surface fluxes are provided in Table~\ref{tab:surfaceflux_LTT1445A} for \LTT\ and Table~\ref{tab:surfaceflux_GJ486} for \GJ. We computed the surface fluxes as $F_{\mathrm{Surf}} = F_{\mathrm{Obs}} \times\ (d/R_{\mathrm{s}})^2 $, where $F_\mathrm{Obs}$ is the integrated observed flux, $d$ is the stellar distance, and $R_\mathrm{s}$ is the stellar radius. Values and uncertainties for $d$ and $R_\mathrm{s}$ for each target are provided in Table~\ref{tab:both_stellar_params}. We propagated the uncertainties in $d$ and $R_\mathrm{s}$ through to the reported surface fluxes; however the resulting flux uncertainties are dominated by fitting the convolved Voigt profiles to the data.
\begin{table*}
\centering
\caption{Measured emission lines from \LTT\ with HST/COS\label{tab:surfaceflux_LTT1445A}}
\begin{tabular}{c|cccccc}
\hline\hline
 & Grating & Total Exposure Time & Line & Line Centers\tsup{a} & log$_{10}$(Surface Flux)\tsup{b} & log$_{10}$($T$)\tsup{a}\\
 &         &   (s)               &      & (\AA)                & (erg cm$^{-2}$ s$^{-1}$) & 
 \\
 \hline
\multirow{9}{*}{\rotatebox[origin=c]{90}{FUV}} & \multirow{6}{*}{G130M} & \multirow{6}{*}{7657} & \multirow{2}{*}{C \textsc{iii} *} & 1174.933, 1175.263, 1175.711,    &  \multirow{2}{*}{$2.99\pm0.12$} &  \multirow{2}{*}{4.9}\\
                    &       &       &                  & 1175.987, 1176.37 &        &     \\
                    &       &       & Si \textsc{iii} *  & 1206.555            & $2.85\pm0.07$ &  4.8\\
                    &       &       & N \textsc{v} *     & 1238.821, 1242.804  & $2.99\pm0.06$ & 5.3\\
                    &       &       & Si \textsc{ii} *   & 1264.738            & $1.99\pm0.29$ & 4.5\\
                    &       &       & C \textsc{ii} *    & 1334.532, 1335.707  & $3.10\pm0.06$ & 4.6\\
                    \cline{2-7}
                    & \multirow{5}{*}{G160M} & \multirow{5}{*}{6830}   & O \textsc{v}    & 1371.296 & $1.73\pm0.66$  & 5.4 \\
                    &       &       & Si \textsc{iv} *   & 1393.755, 1402.772  & $2.95\pm0.14$ & 4.9\\
                    &       &       & O \textsc{iv} & 1401.163 & $1.84\pm0.64$ & 5.1 \\
                    
                    &       &       & C \textsc{iv} * & 1548.187, 1550.775 & $3.61\pm0.09$ & 5.1\\
                    &       &       & He \textsc{ii}   & 1640.474            & $3.20\pm0.19$ & 4.9\\
                    &       &       & Al \textsc{ii}   & 1670.788          & $2.89\pm0.34$ & 4.5\\
\cline{1-7}
\multirow{2}{*}{\rotatebox[origin=c]{90}{NUV}} & \multirow{2}{*}{G230L} & \multirow{2}{*}{540}   & \multirow{2}{*}{Mg \textsc{ii} *}   & \multirow{2}{*}{2796.350, 2803.531} & \multirow{2}{*}{$4.79\pm0.08$} & \multirow{2}{*}{4.2}\\
\\
\hline
\end{tabular}
\tablefoot{Lines marked with an asterisk $*$ are used to compute the DEM function (see Sect.\,\ref{subsec:euv} and Fig.\,\ref{fig:DEM_LTT1445A}). \\
 \tsup{a} Values from CHIANTI database v10.1 \citep{Dere2023}\\
 \tsup{b}Surface fluxes (erg cm$^{-2}$ s$^{-1}$) were calculated by scaling the observed flux by the distance $d$ and stellar radius $R_{\mathrm{s}}$ of \LTT: $F_{\mathrm{Surf}} = F_{\mathrm{Obs}} \times\ (d/R_{\mathrm{s}})^2 $. For multiple lines, we report the combined surface flux.\\
 }
\end{table*}
\begin{table*}
\centering
\caption{Measured emission lines from \GJ\ with HST/COS\label{tab:surfaceflux_GJ486}}
\begin{tabular}{c|cccccc}
\hline\hline
 & Grating & Total Exposure Time & Line & Line Centers\tsup{a} & log$_{10}$(Surface Flux)\tsup{b} & log$_{10}$($T$)\tsup{a}\\
 &         &   (s)               &      & (\AA)                & (erg cm$^{-2}$ s$^{-1}$) & 
 \\
 \hline
\multirow{9}{*}{\rotatebox[origin=c]{90}{FUV}} & \multirow{6}{*}{G130M} & \multirow{6}{*}{7610} & \multirow{2}{*}{C \textsc{iii} *} & 1174.933, 1175.263, 1175.711,    &  \multirow{2}{*}{$3.8\pm0.07$} &  \multirow{2}{*}{4.9}\\
                    &       &       &                  & 1175.987, 1176.37 &        &     \\
                    &       &       & Si \textsc{iii} *  & 1206.555            & $3.04\pm0.08$ &  4.8\\
                    &       &       & N \textsc{v} *     & 1238.821, 1242.804  & $3.27\pm0.07$ & 5.3\\
                    &       &       & Si \textsc{ii} *   & 1264.738            & $2.58\pm0.24$ & 4.5\\
                    &       &       & C \textsc{ii} *    & 1334.532, 1335.707  & $3.71\pm0.06$ & 4.6\\
                    \cline{2-7}
                    & \multirow{5}{*}{G160M} & \multirow{5}{*}{6753}   & O \textsc{v}    & 1371.296 & $1.20\pm0.51$  & 5.4 \\
                    &       &       & Si \textsc{iv} *   & 1393.755, 1402.772  & $2.52\pm0.12$ & 4.9\\
                    &       &       & O \textsc{iv} & 1401.163 & $1.41\pm0.54$ & 5.1 \\
                    
                    &       &       & C \textsc{iv} * & 1548.187, 1550.775 & $3.29\pm0.07$ & 5.1\\
                    &       &       & He \textsc{ii}   & 1640.474          & $2.74\pm0.15$ & 4.9\\
                    &       &       & Al \textsc{ii}  & 1670.788           & $2.6\pm0.23$ & 4.5 \\
\cline{1-7}
\multirow{2}{*}{\rotatebox[origin=c]{90}{NUV}} & \multirow{2}{*}{G230L} & \multirow{2}{*}{552}   & \multirow{2}{*}{Mg \textsc{ii} *}   & \multirow{2}{*}{2796.350, 2803.531} & \multirow{2}{*}{$4.53\pm0.08$} & \multirow{2}{*}{4.2}\\
\\
\hline
\end{tabular}
\tablefoot{Same as Table~\ref{tab:surfaceflux_LTT1445A} but for \GJ. DEM shown in Fig.\,\ref{fig:DEM_GJ486}.
}
\end{table*}

\subsection{Ly\texorpdfstring{$\alpha$}{-alpha}}\label{subsec:lya}
The core of the \Lya\ line is attenuated by resonant scattering of neutral hydrogen in the interstellar medium (ISM); however for bright enough stars it is possible to measure the wings of the \Lya\ profile, and then ``reconstruct'' the intrinsic \Lya\ line \citep{Youngblood2016,Youngblood2021,Youngblood2022}. An alternative method for stars not bright enough to perform the reconstruction is to use known correlations between other UV lines to ``estimate'' the \Lya\ flux \citep{Youngblood2017,Diamond-Lowe2021,Diamond-Lowe2022}. \LTT\ and \GJ\ are both bright enough to do a \Lya\ reconstruction, which we used in the panchromatic spectra. We also compared the reconstructed \Lya\ flux to the estimated \Lya\ flux from UV-UV line correlations. 

\begin{figure}
\resizebox{\hsize}{!}{\includegraphics{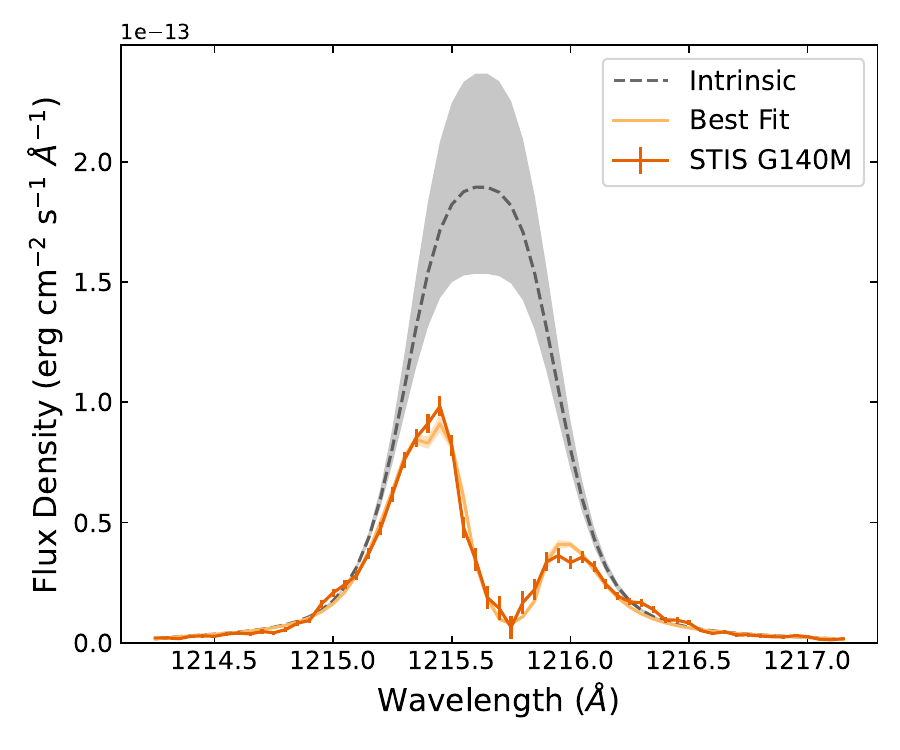}}
\caption{Reconstruction of the intrinsic \Lya\ flux of \LTT\ (gray) using the best fit (light orange) to the measured wings of the \Lya\ profile observed with STIS/G140M \citep[orange 1$\sigma$ error bars;][]{Youngblood2016}.
} \label{fig:LyaRecon_LTT1445A}%
\end{figure}

\begin{figure}
\resizebox{\hsize}{!}{\includegraphics{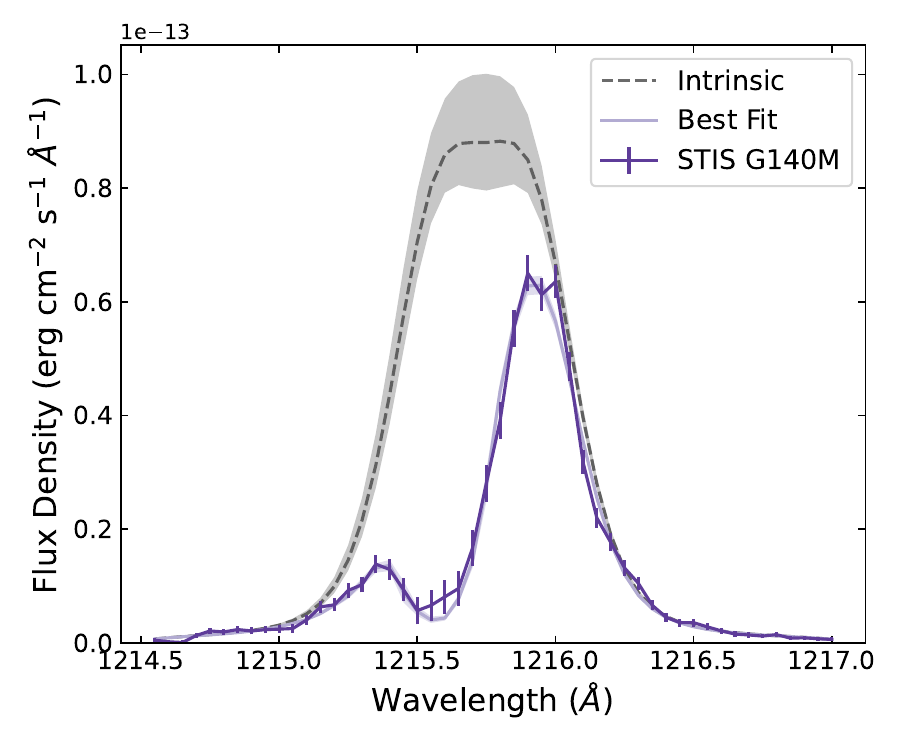}}
\caption{Same as Fig.\,\ref{fig:LyaRecon_LTT1445A} but for \GJ. 
} \label{fig:LyaRecon_GJ486}%
\end{figure}

We combined three STIS/G140M measurements and took a weighted average to build up S/N for the \Lya\ reconstruction (Figs.\,\ref{fig:LyaRecon_LTT1445A} and~\ref{fig:LyaRecon_GJ486}). In the case of \LTT, one of the three observations, from GO 16722 Visit 03, appeared to contain about 1.5$\times$ the flux of the other two observations; we posit that this additional flux may be associated with the flare that preceded the observation by 48 hours  (Fig.\,\ref{fig:timeseries_LTT1445A}). A further discussion of prolonged stellar activity can be found in Sect.\,\ref{subsec:stellar_activity}.

\begin{figure}
\resizebox{\hsize}{!}{\includegraphics{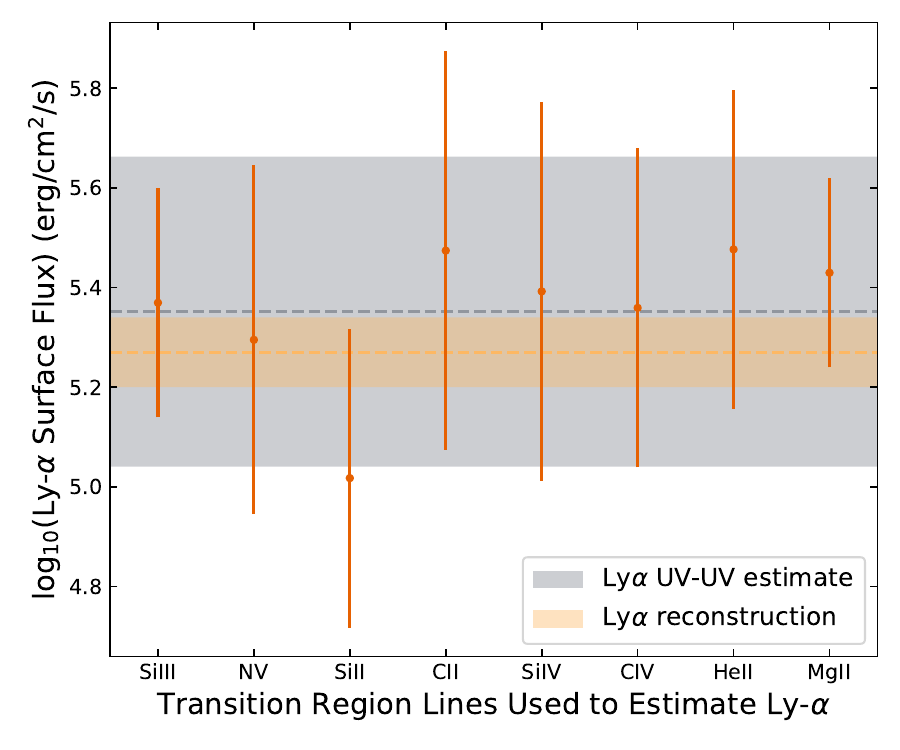}}
\caption{Deriving the \Lya\ flux of \LTT\ using a reconstruction from the wings of the \Lya\ profile observed with the STIS/G140M grating \citep[orange 1$\sigma$ band;][]{Youngblood2016} and using UV--UV line correlations with other measured UV lines with the COS instrument \citep[orange points with 1$\sigma$ errors, and the average 1$\sigma$ gray band;][]{Youngblood2017,Diamond-Lowe2021,Diamond-Lowe2022}. In the case of \LTT,\ we find that individual transition region lines agree with each other, and with the reconstructed value.
} \label{fig:Lya_LTT1445A}%
\end{figure}

For comparison, we estimated \Lya\ flux for \LTT\ from UV--UV line correlations found in the MUSCLES sample \citep{Youngblood2017}. In addition to demonstrating the ability of the UV--UV correlations to recover the data-driven \Lya\ reconstruction, this method additionally provides a useful comparison of the UV flux from transition region lines observed with different COS gratings, which cannot be used simultaneously. As demonstrated in Fig.\,\ref{fig:Lya_LTT1445A}, the UV--UV correlation values for \Lya\ estimated from individual transition region lines agree well with each other, and the resulting average \Lya\ estimate agrees well with the reconstructed \Lya\ value. Similar to previous work \citep[e.g.,][]{Diamond-Lowe2022}, we conservatively estimated the uncertainty of the individual \Lya\ estimates using the rms of the correlations from \citet{Youngblood2017}, and took the final estimate as the mean value and the mean uncertainty. 

\begin{figure}
\resizebox{\hsize}{!}{\includegraphics{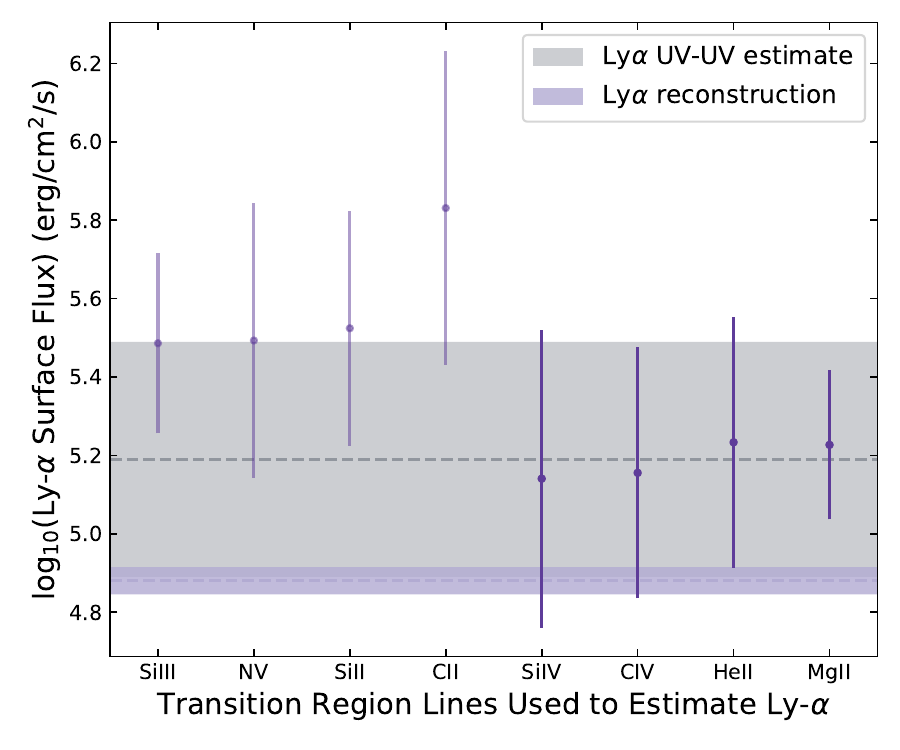}}
\caption{Same as Fig.\,\ref{fig:Lya_LTT1445A} but for \GJ. In this case, we find a discrepancy between the lines observed with COS/G130M and those observed with COS/G160M and COS/G230L. It is possible that this discrepancy is due to the large flares observed during the COS/G130M observations. We only use lines observed with COS/G160M and COS/G230L (darker purple points with 1$\sigma$ errors) to estimate the \Lya\ flux with the UV--UV correlation method. The \Lya\ values from the reconstruction and the UV--UV estimation agree to within 2$\sigma$. 
} \label{fig:Lya_GJ486}%
\end{figure}

In the case of \GJ, there is a potential discrepancy between the UV--UV correlation values for \Lya\ from transition region lines measured with the COS/G130M grating and those measured with the COS/G160M and G230L gratings (Fig.\,\ref{fig:Lya_GJ486}). During observations with the COS/G130M grating we detected two large flares (Fig.\,\ref{fig:timeseries_GJ486}). These flares are removed from the data and analyzed separately in Sect.\,\ref{sec:flare}, but the remaining ``quiescent'' data may still represent an elevated activity state (more on this in Sect.\,\ref{subsec:stellar_activity}). The discrepancy between ``quiescent'' UV estimates of \Lya\ suggests that we have not actually observed \GJ\ in true quiescence with the COS/G130M grating. For the purposes of comparison with the \Lya\ reconstruction, we only used the lines observed with COS/G160M and COS/G230L, which were observed about a week after the COS/G130M observations. The UV-UV correlation values for \Lya\ estimated from the COS/G160M and G230L lines agree with the \Lya\ value at the 2$\sigma$ level.

The reconstructed \Lya\ profiles for \LTT\ and \GJ\ are included in the final panchromatic spectrum. Properties derived from the reconstruction are provided in Table~\ref{tab:Lya_both}, along with a comparison to the UV--UV correlation method. 
\begin{table*}
\centering
\caption{\Lya\ values for \LTT\ and \GJ \label{tab:Lya_both}}
\begin{tabular}{l|cc}
\hline\hline
 &                LTT 1445A & GJ 486                \\ 
 \hline
 &                \multicolumn{2}{c}{Reconstruction from STIS/G140M} \\
\hline
log\tsub{10}(Surface Flux) (erg/cm\tsup{2}/s)  & $5.27^{+0.08}_{-0.06}$ & $4.88^{+0.04}_{-0.03}$ \\
log\tsub{10}(N(H \textsc{i})) (cm\tsup{-2}) & $17.8\pm0.1$ & $17.8\pm0.1$ \\
H \textsc{i} radial velocity (km/s)    & $13.2^{+3.4}_{-2.8}$       & $-18.1^{+1.3}_{-2.0}$ \\
\hline
 &                \multicolumn{2}{c}{Estimate from UV-UV Correlations} \\
 \hline
log\tsub{10}(Surface Flux) from Si \textsc{iii} (erg/cm\tsup{2}/s) & $5.37\pm0.23$ & $5.48\pm0.23$ \\
log\tsub{10}(Surface Flux) from N \textsc{v} (erg/cm\tsup{2}/s)    & $5.30\pm0.35$ & $5.49\pm0.35$   \\
log\tsub{10}(Surface Flux) from Si \textsc{ii} (erg/cm\tsup{2}/s)  & $5.02\pm0.30$ & $5.52\pm0.30$   \\
log\tsub{10}(Surface Flux) from C \textsc{ii} (erg/cm\tsup{2}/s)   & $5.48\pm0.40$ & $5.84\pm0.40$   \\
log\tsub{10}(Surface Flux) from Si \textsc{iv} (erg/cm\tsup{2}/s)  & $5.39\pm0.38$ & $5.14\pm0.38$   \\
log\tsub{10}(Surface Flux) from C \textsc{iv} (erg/cm\tsup{2}/s)   & $5.36\pm0.32$ & $5.15\pm0.32$   \\
log\tsub{10}(Surface Flux) from He \textsc{ii} (erg/cm\tsup{2}/s)  & $5.47\pm0.32$ & $5.23\pm0.32$   \\
log\tsub{10}(Surface Flux) from Mg \textsc{ii} (erg/cm\tsup{2}/s)  & $5.43\pm0.19$ & $5.22\pm0.19$   \\
\hline
log\tsub{10}(Surface Flux) mean (erg/cm\tsup{2}/s)  & $5.35\pm0.31$ & $5.19\pm0.30$\tsup{a} \\
\hline
\end{tabular}
\tablefoot{Log surface fluxes calculated using distances and stellar radii from Table~\ref{tab:both_stellar_params}.\\
\tsup{a} 
 For \GJ\ we find that lines measured with the COS/G130L grating give systematically higher estimates of the \Lya\ flux. Given that two large flares were removed from the data taken with this grating, it is possible that the transition region lines we measure with COS/G130L are not completely representative of a quiescent state. We therefore report the \Lya\ estimate from the UV--UV correlations only with lines in the G160M and G230L gratings for \LTT.}
\end{table*}

\subsection{X-ray}
\label{ssec:xraySpec}

Detecting X-ray flux for most inactive mid-to-late M dwarfs is challenging as their apparent brightness at these wavelengths is typically faint. \LTT\ and \GJ\ are close enough that X-ray detections are possible. We used \textit{Chandra} to measure X-ray flux from \LTT\ and both \textit{Chandra} and \textit{XMM-Newton} to measure X-ray flux from \GJ. We detected X-rays flares for both stars, which are discussed in more detail in Sect.\,\ref{sec:flare}. Similar to the UV data, we removed the flare points from the time-series X-ray data to measure quiescent X-ray flux levels. The quiescent X-ray flux measurements were used to construct the panchromatic spectra and inform the DEM estimate of the EUV (Sect.\,\ref{subsec:euv}). Quiescent and flare values from the X-ray observations are provided in Tables~\ref{tab:Xray_LTT1445A} and~\ref{tab:Xray_GJ486} for \LTT\ and \GJ, respectively.

\begin{figure*}
\sidecaption
\includegraphics[width=12cm]{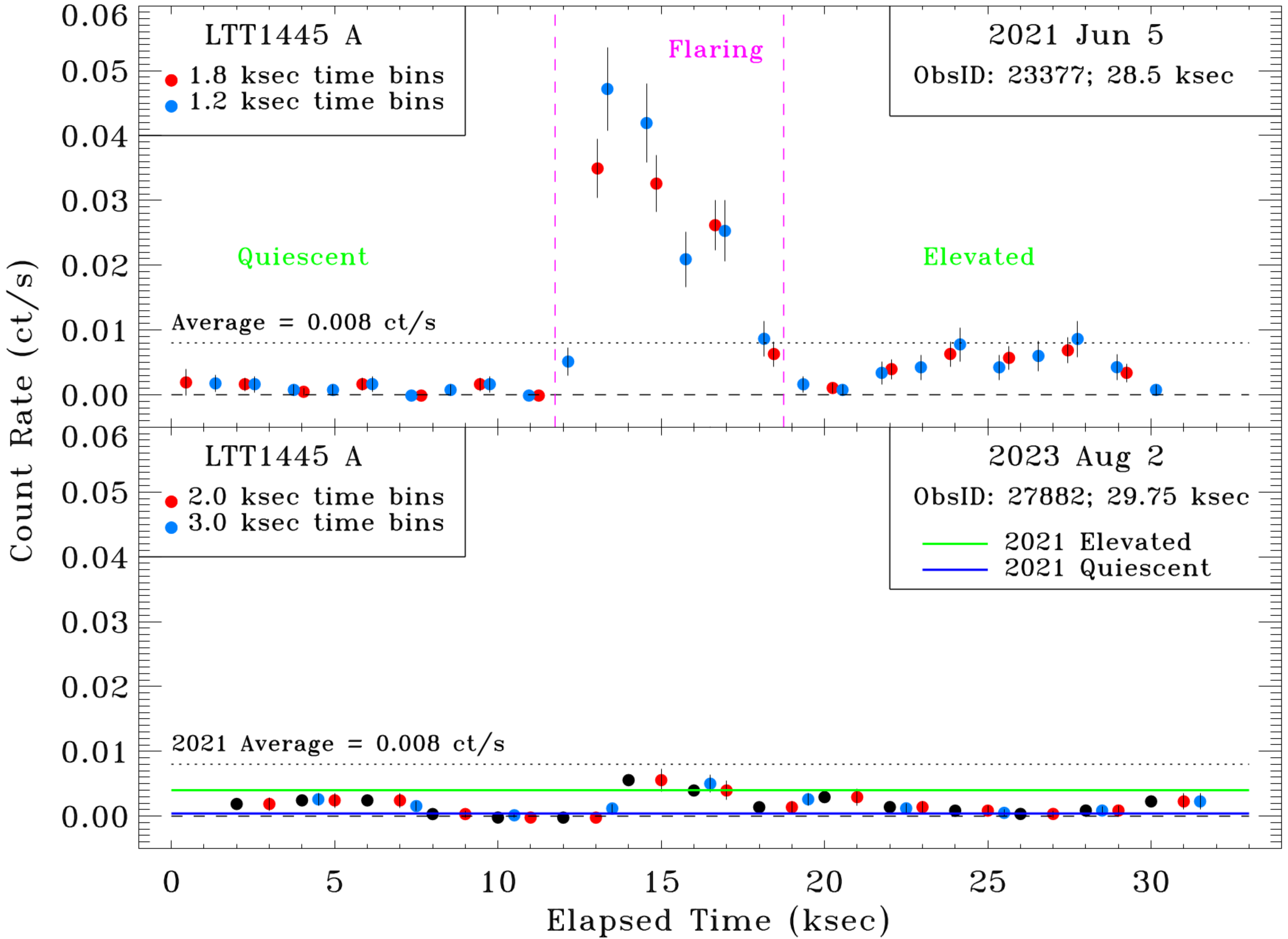}
\caption{X-ray observations of \LTT\ using \textit{Chandra} ACIS-S4. Observations from 2021 (top panel) were already reported in \citet{Brown2022}. The 2021 observations distinctly show a flare. We use the observations from 2023 (Obs ID 27882, PI Howard) as representative of the typical X-ray state of \LTT\ (bottom panel).}
\label{fig:Xray_LTT1445A}%
\end{figure*}

For \LTT\ we used \textit{Chandra} observations published in \citet{Brown2022}, as well as a new set of observations presented here (Fig.\,\ref{fig:Xray_LTT1445A}, Table~\ref{tab:Xray_LTT1445A}). The first set of observations, obtained in 2021, distinctly show a flare. An extensive analysis of this flare is reported in \citet{Brown2022} and we do not repeat the process here. We observed \LTT\ again on 2023 Aug 2 (Obs ID 27882, PI Howard) and did not detect a flare. Estimation of X-ray flux parameters involves spectral fitting using the XSPEC v12.12 software \citep{Arnaud1996}. Count rates were converted to flux values in erg s$^{-1}$ cm$^{-2}$ using the best-fit single-temperature VAPEC spectrum for an optically thin coronal plasma, assuming a hydrogen column density of 10$^{19}$ cm$^{-2}$, subsolar metallicities, and stellar distance provided in Table~\ref{tab:both_stellar_params}. A full description of how we chose the subsolar abundances can be found in Sect.\ 4.3 of \citet{Brown2022}. We take the 2023 observations as representative of \LTT's typical X-ray state, and divided the flux into three broad spectral bins for use in the DEM (Sect.\,\ref{subsec:euv}). We used only the 2023 observations in the \GJ\ panchromatic spectrum.

\begin{figure}
\resizebox{\hsize}{!}{\includegraphics{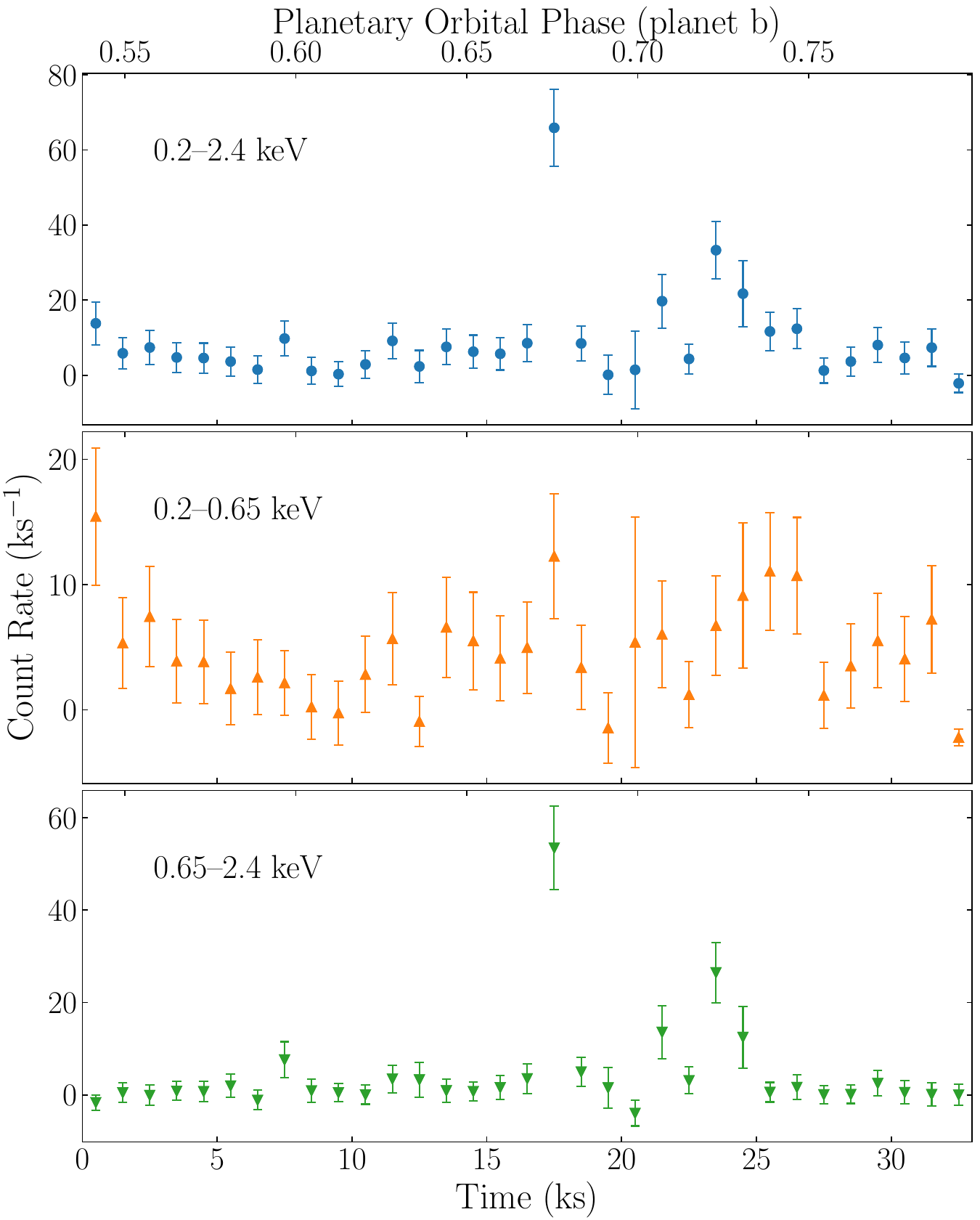}}
\caption{\textit{XMM-Newton} EPIC light curve of \GJ\ from the 2021 Dec 23 observation. The count rates are co-added across the three EPIC cameras (pn, MOS1 and MOS2). The three panels show the count rate in three bands: 0.2--2.4\,keV (top), 0.2--0.65\,keV (middle), and 0.65--2.4\,keV (bottom).}
\label{fig:Xray_GJ486}%
\end{figure}

For \GJ\ we were able to use \textit{XMM-Newton}, which has better sensitivity at lower energies. We observed \GJ\ for 32\,ks on 2021 Dec 23 with the European Photon Imaging Camera (EPIC). A few short periods of high background flaring associated with energetic Solar protons were excluded \citep{Walsh2014} using the standard processes\footnote{As outlined in the SAS threads: \url{https://www.cosmos.esa.int/web/xmm-newton/sas-threads}}. Our extracted light curves of \GJ\ are displayed in Fig.\,\ref{fig:Xray_GJ486}, with results for three different energy bands: ``full'' (0.2--2.4\,keV), ``soft'' (0.2--0.65\,keV), and ``hard'' (0.65--2.4\,keV). These light curves show evidence of flares, which we describe in more detail in Sect.\,\ref{ssec:xrayFlare}.

We extracted separate EPIC-pn spectra for the quiescent epochs of the observation and the biggest flare, which occurred halfway through the time series. The S/N of the second peak was too low to warrant fitting, but this section of the observation was excluded from the quiescent spectrum. We fit the data using a two temperature APEC model, where the temperatures were forced to be the same across the quiescent and flaring spectra, due to the latter only containing five spectral bins. However their respective normalizations are allowed to vary, such that the substantial change in the spectral shape can be accounted for in the fit. We again use the same subsolar abundances from \citet{Brown2022} for these fits, noting that our best fit fixes the Fe abundance, as this parameter becomes unconstrained when allowed to vary. We used C-statistics when performing the fit \citep{Cash1979}. From the best-fit model, we calculated the flux in several broad energy ranges for use in reconstructing the DEM (Sect.\,\ref{subsec:euv}).

We also have observations of \GJ\ using the \textit{Chandra} HRC-I detector obtained on 2023 April 4, July 13, and July 15. These observations appeared to show quiescent emission and the corresponding X-ray luminosity, and we estimated the emission measures assuming a coronal temperature of 0.19 keV (2.2 MK)\footnote{See \citet{Brown2023} for a discussion on quiescent M dwarf coronal temperatures}, because the HRC-I has minimal energy resolution and does not provide a temperature directly.
\begin{table*}
\centering
\caption{X-ray source properties for \LTT \label{tab:Xray_LTT1445A}}
\begin{tabular}{l|ccc}
\hline\hline
 &                \multicolumn{3}{c}{\textit{Chandra} ACIS-S4 } \\
  &                         2021 & 2021                     & 2023              \\ 
 &                  Quiescent  &  Flare              & All             \\ 
\hline
ObsID (PI) & \multicolumn{2}{c}{23377 (Brown)} & 27882 (Howard)\\
Exp.\ time (ks)     & 12.2        & 6.66        & 29.75\\
Source counts (ct)  & 4.9         & 177        & 50\\
Count rate (ct/ks)  & 0.4 ± 0.2   & 26.6 ± 2.0   &  1.68 ± 0.25   \\
X-ray Flux (10\tsup{-13} erg/cm\tsup{2}/s) & 0.066 ± 0.033 & 3.61 ± 0.27  & 0.51 ± 0.08  \\
Characteristic Temperature (keV)& 0.5\tsup{a} &1.02$\pm$0.10 & 0.55$\pm$0.20 \\
log\tsub{10}(L$_X$) (erg/s) & 25.57$^{+0.18}_{-0.30}$ & 27.31 ± 0.03  &  26.46 ± 0.04  \\
\hline
\end{tabular}
\tablefoot{Values from 2021 reproduced from \citet{Brown2022}. Log fluxes calculated assuming stellar distance provided in Table~\ref{tab:both_stellar_params}. \\
\tsup{a} We assume this temperature from the measurement from the elevated flux post-flare; it is in agreement with the 2023 observations.
}
\end{table*}
\begin{table*}
\centering
\caption{X-ray source properties for \GJ \label{tab:Xray_GJ486}}
\begin{tabular}{l|ccc}
\hline\hline
 &      \multicolumn{2}{c}{\textit{XMM-Newton} EPIC-pn} &     \textit{Chandra} HRC-I \\
 &               Quiescent &  Flare 1  & All             \\ 
\hline
Energy range (keV)     & \multicolumn{2}{c}{0.2--2.4} & 0.1--10\\
Wavelength range (\AA) & \multicolumn{2}{c}{5.2--62} & 1.2--124 \\
Obs ID (PI)             & \multicolumn{2}{c}{0892010101 (Co-PI King)} & 26210/27799/27942 (Youngblood)\\
Exp.\ time (ks)        & 26.02     &    1.73       & 31.54  \\
Source counts (ct)     & 109±15    &    33.2±6.4   & 55.8 ± 7.8  \\
Count rate (ct/ks)     & 4.18±0.57 &    19.2±3.7   & 1.77 ± 0.25  \\
X-ray Flux (10\tsup{-14} erg/cm\tsup{2}/s) & $0.80^{+0.68}_{-1.64}$     &    $4.35^{+0.61}_{-0.83}$         & 1.88 ± 0.27 \\
Temperature 1 (keV) & \multicolumn{2}{c}{$0.115^{+0.022}_{0.021}$} & 0.19 (assumed)\\
Temperature 2 (keV) & \multicolumn{2}{c}{$0.783^{+0.097}_{0.099}$} & n/a\\
log\tsub{10}(L$_X$) (erg/s)  &  $25.797^{+0.035}_{-0.099}$ &    $26.531^{0.057}_{0.092}$       & 26.17 ± 0.07    \\
\hline
\end{tabular}
\tablefoot{The ``Flare 1'' spectrum also includes the underlying quiescent emission at that epoch. The spectrum for the second flare had insufficient signal to warrant fitting. Flux is the unabsorbed flux at Earth. Log luminosities calculated assuming stellar distance provided in Table~\ref{tab:both_stellar_params}.
 }
\end{table*}

\subsection{Extreme-ultraviolet}\label{subsec:euv}

There is no currently operating observatory that can detect EUV (100--912\AA) flux from \LTT\ and \GJ. Energetic photons approaching 912\AA\ are increasingly absorbed by neutral hydrogen in the ISM, making it nearly impossible to detect photons at 912\AA\ for stars other than the Sun. The Extreme-Ultraviolet Explorer \citep[EUVE;][]{Craig1997}, which functioned from 1992--2001, could measure EUV flux from nearby, highly energetic stars, but neither \LTT\ nor \GJ\ were observed by this mission. Instead, in order to estimate the EUV flux from \LTT\ and \GJ, we leveraged the measured transition lines we detect in the UV, as well as the X-ray measurements, in order to construct a DEM function for our targets.  

The DEM method takes advantage of the fact that a smoothly varying function with respect to temperature can be fit to the DEMs calculated for observed emission lines in the data; the fitted function can then be used to back out the flux we should measure from emission lines where no data are available, such as those at EUV wavelengths. There are many examples of DEM applications in stellar astrophysics \citep[e.g.,][]{Kashyap&Drake1998,Sanz-Forcada2003}. Here we followed the application of the DEM method to cool dwarf stars \citep{Duvvuri2021}, with several examples now available \citep[e.g.,][]{Diamond-Lowe2021,Diamond-Lowe2022,Wilson2021,Feinstein2022,Duvvuri2023}. The DEM method relies on some assumptions, for example that each temperature component of the stellar atmosphere can be treated as an optically thin plasma in collisional ionization equilibrium. These assumptions limit the DEM method, but without another way to access the EUV spectra of inactive M dwarfs, the DEM method is the current state-of-the-art for estimating EUV flux for our targets. We note that it is also possible to estimate the EUV flux by scaling from the \Lya\ line \citep{Linsky2014} or the Si \textsc{iv} and N \textsc{v} lines \citep{France2018}. \citet{Diamond-Lowe2021} found agreement between these scaling methods and the DEM method for the inactive M dwarf LHS 3844. Here we opt for the DEM method because it utilizes measured flux across the UV and X-ray.

\begin{figure*}
\sidecaption
\includegraphics[width=12cm]{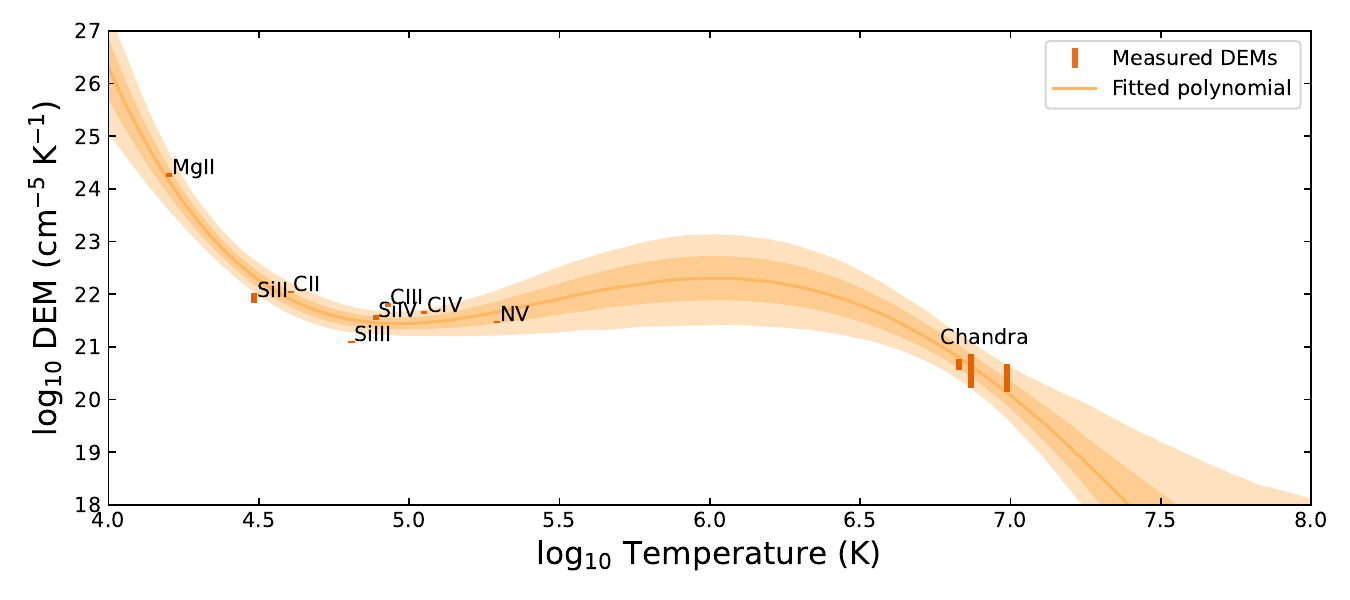}
\caption{Differential emission measures (DEM) with 1$\sigma$ uncertainties derived from measured UV flux from HST/COS and X-ray flux from \textit{Chandra} ACIS of \LTT. We show the fifth-order Chebyshev polynomial best fit to the local DEMs along with shaded regions representing the 1$\sigma$ and 2$\sigma$ uncertainties for the fit.}
          \label{fig:DEM_LTT1445A}%
\end{figure*}

\begin{figure*}
\sidecaption
\includegraphics[width=12cm]{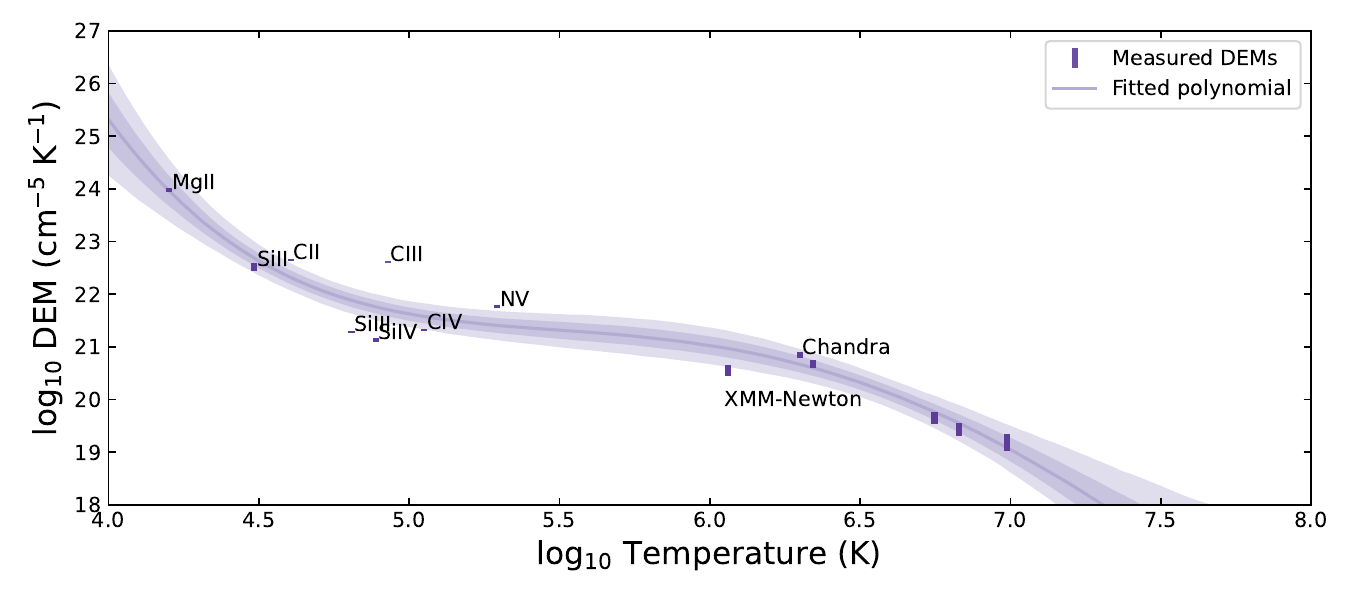}
\caption{Same as Fig.\,\ref{fig:DEM_LTT1445A} but for \GJ. All X-ray DEMs are derived from \textit{XMM-Newton} observations, except for the single point from \textit{Chandra} HRC-I.}
          \label{fig:DEM_GJ486}%
\end{figure*}

Following methods outlined in \citet{Duvvuri2021} and improving upon code developed in \citet{Diamond-Lowe2021} and \citet{Diamond-Lowe2022}, we constructed a DEM for \LTT\ and \GJ. For each observed UV line and X-ray band we calculated a ``local DEM'' that is the local average DEM value required to reproduce the observed flux given the method's assumptions. We used the CHIANTI atomic database v10.1 to retrieve the maximum formation temperature and emissivity contribution functions for each ion \citep{Dere1997,Dere2023}. For the X-ray bands we used the CHIANTI database to find every ion that emits in each band, and summed their emissivity contribution functions. We used the peak of this summed function to determine a peak formation temperature. The exception is the \textit{Chandra} HRC-I observation for \GJ, which did not have any energy resolution, so we took the assumed coronal temperature of 2.2MK as the peak formation temperature. We fit a fifth-order Chebyshev polynomial to the local DEMs, and used the \texttt{dynesty} dynamic nested sampler to explore the parameter space within the prior bounds. We set priors on the Chebyshev polynomial coefficients as prescribed in \citet{Duvvuri2021}, as well as included the free parameter $s$ to characterize unknown systematic uncertainties. The resulting DEM functions for \LTT\ and \GJ\ are shown in Figs.\,\ref{fig:DEM_LTT1445A} and~\ref{fig:DEM_GJ486}, respectively. 

We note that because we do not have any measured EUV constraints for our targets, it is likely that the fitted DEM will over-predict the EUV flux \citep{DelZanna2002}. This over-prediction is nonuniform and therefore cannot be corrected with a simple scaling. \citet{Duvvuri2021} found an overestimation by a factor of $\sim$5 for stars where they could compare results with and without measured EUV DEMs. In the case of \LTT\ and \GJ\ we do not have direct flux measurements at EUV wavelengths, so this error is unavoidable and we cannot know how much the EUV flux is over-estimated in our analysis. However, we do note that by sampling the parameters that describe the DEM function, a lack of data increases the resulting uncertainty. This is most apparent in the case of \LTT\ where we do not have as much temperature-coverage for the DEM function in the range of 6.0 $<$ log\tsub{10}($T$) $<$ 6.5. The resulting \LTT\ spectrum is more likely to suffer from the over-prediction noted by \citet{DelZanna2002}; however the fitted DEM also has a greater uncertainty, and so the EUV flux error of the resulting spectrum encompasses a factor of 5 at the 1$\sigma$ level in almost all EUV flux bins, and at the 2$\sigma$ level in all EUV flux bins.

\subsection{Optical and infrared}\label{subsec:optical_ir}

To fill in the optical and infrared end of the panchromatic spectra we used BT-SETTL (CIFIST) models that cover the optical to infrared part of the spectrum out to 20\,$\mu$m \citep{Allard2003,Allard2007,Allard2011,Allard2012,Allard2013,Barber2006,Caffau2011} accessed from the Spanish Virtual Observatory database repository\footnote{\href{http://svo2.cab.inta-csic.es/theory/main/}{http://svo2.cab.inta-csic.es/theory/main/}} \citep{Bayo2008}. The grid of stellar spectra were interpolated to the published values of effective temperature $T_\text{eff}$ and surface gravity log($g$) for \LTT\ \citep{Winters2019,Winters2022} and \GJ\ \citep{Trifonov2021,Passegger2019}.

\begin{figure}
\resizebox{\hsize}{!}{\includegraphics{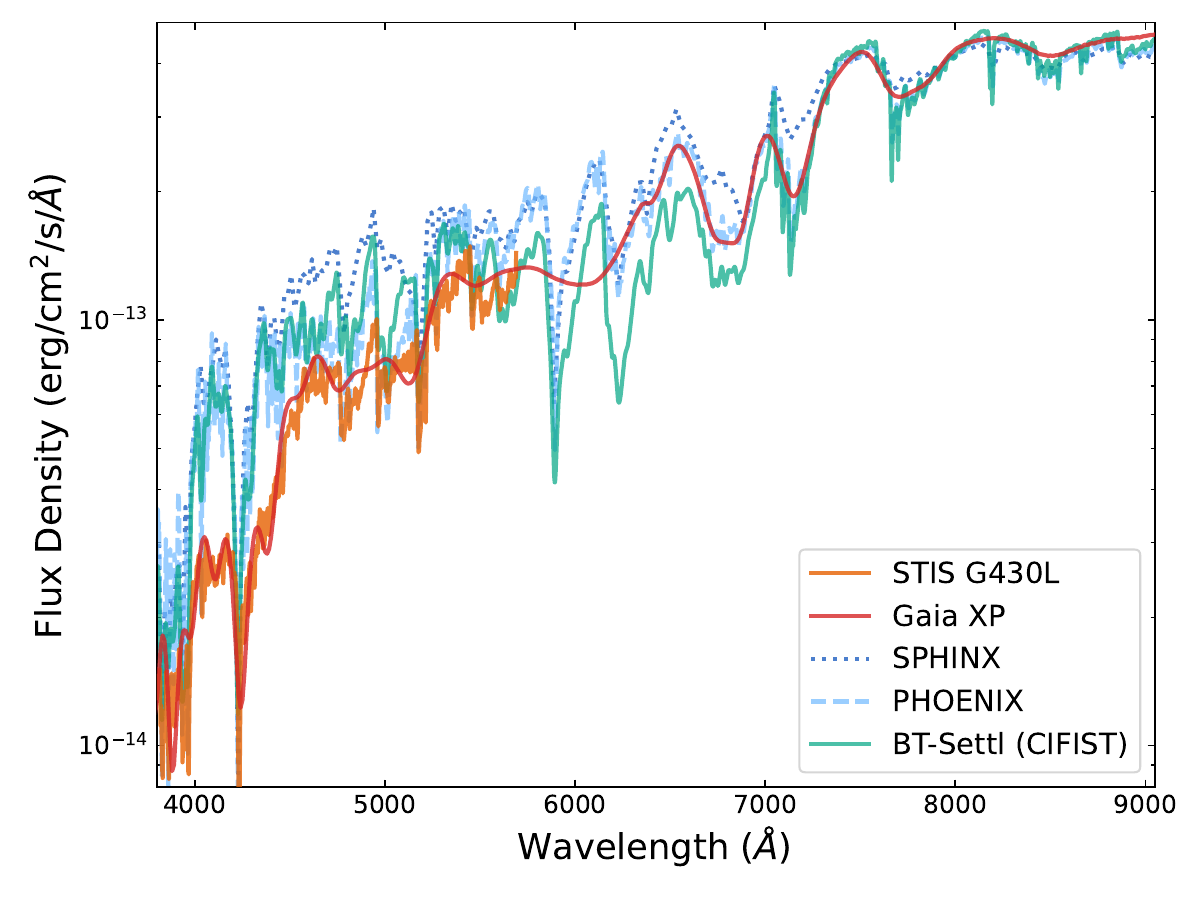}}
\caption{Comparison of flux measured with HST/STIS and the G430L grating, \textit{Gaia} DR3 spectra, and model spectra from SPHINX \citep{Iyer2023}, PHOENIX \citep{Husser2013}, and BT-Settl \citep{Caffau2011} interpolated to the published parameters of \LTT. No relative scaling was performed. The STIS/G430L and \textit{Gaia} XP data agree where they overlap. We use the BT-Settl model in the panchromatic spectrum. 
} \label{fig:comp_spectra_model_LTT1445A}%
\end{figure}

Blue-ward of $\sim$6000\,\AA\ model spectra of M dwarfs do not accurately reproduce measured stellar flux \citep[e.g.,][]{Fontenla2016}. To determine how reliably we can append a stellar model to the HST data, we compared the STIS/G430L observations with spectral data from \textit{Gaia} DR3 BP/RP\footnote{\href{https://gaia-dpci.github.io/GaiaXPy-website/}{https://gaia-dpci.github.io/GaiaXPy-website/}\\ DOI v2.1.0: 10.5281/zenodo.8239995} \citep{GaiaDR32023}, as well as the interpolated stellar model. The STIS/G430L and \textit{Gaia} data agreed where they overlap in wavelength. The interpolated BT-Settl (CIFIST) model broadly agreed with the \textit{Gaia} spectra down to $\sim$5500\AA\, but at shorter wavelengths the model over-predicted the amount of flux measured by HST/STIS G430L and \textit{Gaia} DR3 (Fig.\,\ref{fig:comp_spectra_model_LTT1445A}). 

We also compared interpolated stellar spectra from PHOENIX \citep{Husser2013} and SPHINX \citep{Iyer2023} models, which also over-predicted the stellar flux at short wavelengths. This over-prediction of flux by stellar models is one reason why having measurements from STIS/G430L out to 5700\AA\ is so valuable, especially for M dwarfs. We chose to use the BT-Settl (CIFIST) models because the PHOENIX models do not extend far enough into the infrared and the SPHINX models have less agreement with the \textit{Gaia} and HST data. To complete our panchromatic spectra, we append BT-Settl (CIFIST) models interpolated to the parameters of \LTT\ and \GJ\ to the red end of the HST spectra.
\subsection{Putting it all together}

\begin{figure*}
\sidecaption
\includegraphics[width=12cm]{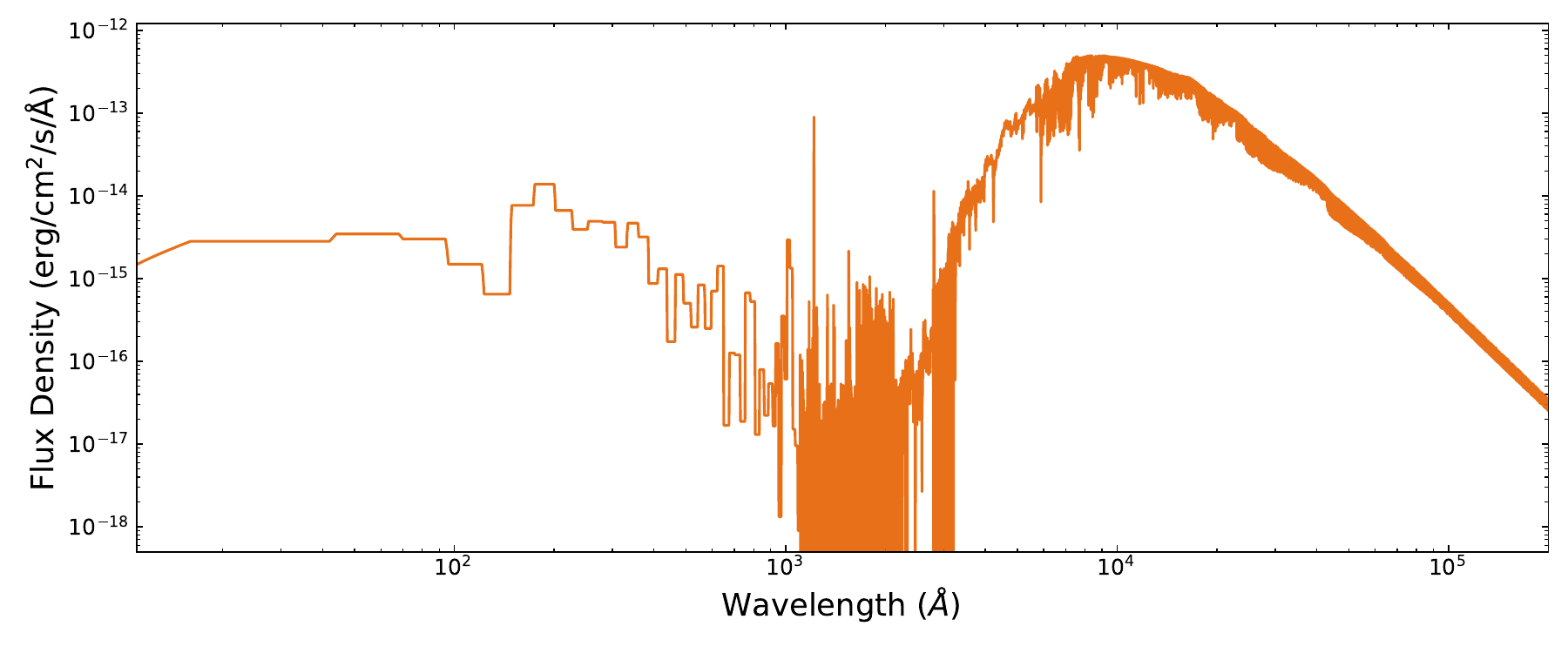}
\caption{Panchromatic spectrum of \LTT\ from 1\,\AA--20\,$\mu$m constructed from empirical data in the X-ray, UV, and optical (HST/COS and STIS, \textit{Chandra}), estimates of the EUV (using the DEM method), a \Lya\ line reconstruction, and models of the optical and infrared (BT-Settl).}
          \label{fig:panchromaticspec_LTT1445A}%
\end{figure*}

\begin{figure*}
\sidecaption
\includegraphics[width=12cm]{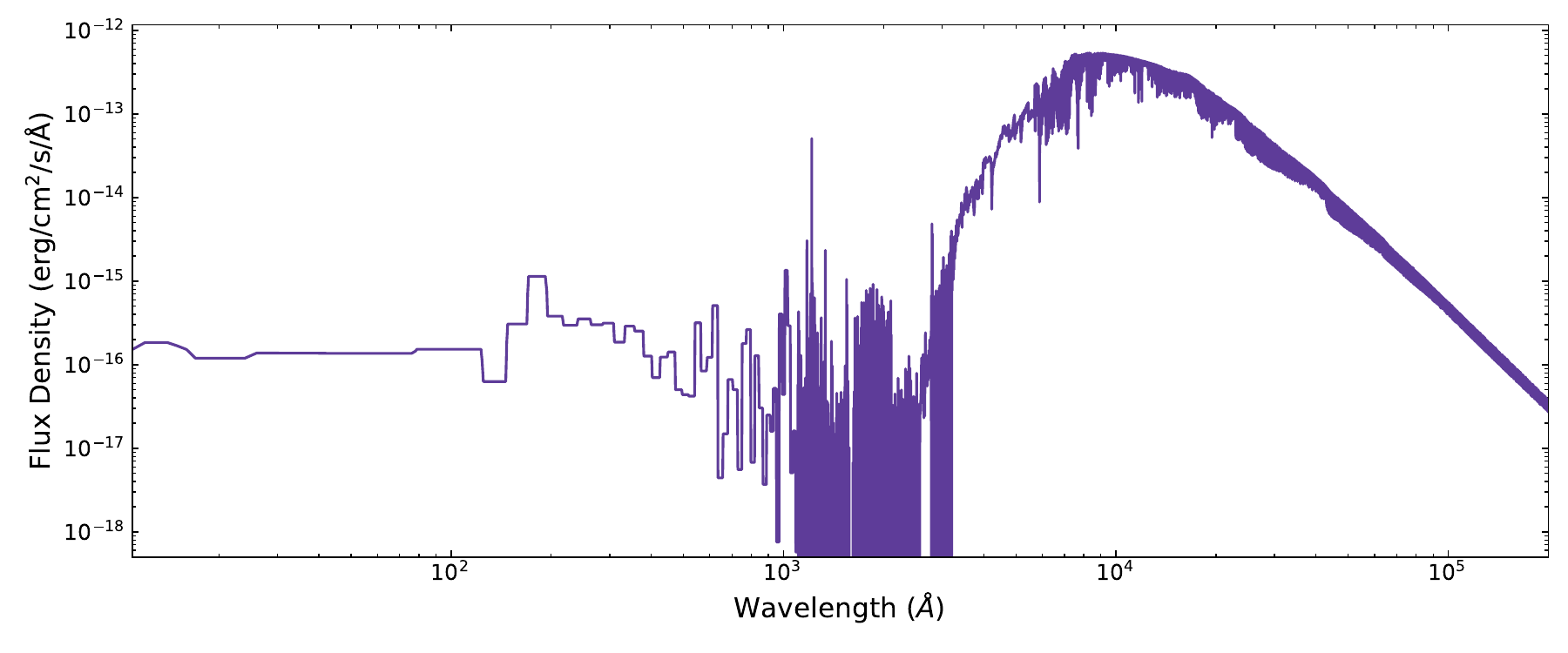}
\caption{Panchromatic spectrum of \GJ\ from 1\,\AA--20\,$\mu$m constructed from empirical data in the X-ray, UV, and optical (HST/COS and STIS, \textit{Chandra}, \textit{XMM-Newton}), estimates of the EUV (using the DEM method), a \Lya\ reconstruction, and models of the optical and infrared (BT-Settl).}
          \label{fig:panchromaticspec_GJ486}%
\end{figure*}

For both \LTT\ and \GJ\ we provide panchromatic spectra representing the quiescent state of each star. These spectra range from 1\,\AA--20\,$\mu$m and are available on the \texttt{mstarpanspec} high level science product (HLSP) page of the MAST archive\footnote{\href{https://archive.stsci.edu/hlsp/mstarpanspec}{https://archive.stsci.edu/hlsp/mstarpanspec}}. As with previous works, the panchromatic spectra are available in four different versions:
\begin{itemize}
    \itemsep0em 
    \item Variable resolution reflecting different instrument and model resolutions
    \item Constant resolution binned to 1\,\AA
    \item Variable resolution, adaptively binned to remove negative flux
    \item Constant resolution at 1\,\AA, adaptively binned to remove negative flux
\end{itemize}
Characteristic spectral properties for \LTT\ and \GJ\ are provided in Table~\ref{tab:spec_properties}, and panchromatic spectra are shown in Figs.\,\ref{fig:panchromaticspec_LTT1445A} and ~\ref{fig:panchromaticspec_GJ486}.
\begin{table*}
\centering
\caption{Derived spectrum values \label{tab:spec_properties}}
\begin{tabular}{lc|cc|c}
\hline
\hline
& Wavelength & \multicolumn{2}{c|}{\LTT} & \GJ \\
\hline
log\tsub{10}($L_{\text{Bol}}$) & 1\AA--20$\mu$m & \multicolumn{2}{c|}{31.49} & 31.66 \\
$f$(XUV) & 1--912\AA     & \multicolumn{2}{c|}{-3.45}                        & -4.55 \\
$f$(FUV) & 912--1700\AA  & \multicolumn{2}{c|}{-4.33}                        & -4.64 \\
$f$(XUV\tsub{age}) & 5--1700 & \multicolumn{2}{c|}{-3.40}                    & -4.30 \\
$f$(NUV) & 1700--3200\AA & \multicolumn{2}{c|}{-4.20}                        & -4.41 \\
FUV/NUV & ---            & \multicolumn{2}{c|}{0.75}                         & 0.58 \\
\Lya/FUV & ---           & \multicolumn{2}{c|}{0.57}                         & 0.49 \\
\hline
         &            & \LTT b                              & \LTT c           & \GJ b \\
\hline
F$_{\mathrm{XUV,p}}$ (erg/cm\tsup{2}/s) & 1--912\AA & $2.67\times10^3$ & $5.49\times10^3$ &  $1.50\times10^3$ \\

\hline
\end{tabular}
\tablefoot{We compute $f$(band) = log$_{10}$($L_{\mathrm{band}}/L_{\mathrm{Bol}}$), where $L_\text{Bol}$ is in units of erg s\tsup{-1}. The value F$_{\mathrm{XUV,p}}$ is the integrated XUV flux (1--912\AA) received by each planet, assuming orbital distance $a$ from Table~\ref{tab:both_stellar_params}. We provide the value $f$(XUV\tsub{age}) becuase it is used to calculate stellar ages in Sect.\,\ref{subsec:starage} based on the XUV activity--age relationship from \citet{Engle2024}.
}
\end{table*}
\section{Flare analysis}\label{sec:flare}

Both \LTT\ and \GJ\ exhibit flares in the UV and X-ray observations. These data are not simultaneous, nor do they cover the same wavelengths, so we cannot determine broadband flare characteristics \citep[e.g., flare temperatures;][]{Berger2023,Jackman2023}. We also cannot compare the flare energies between the UV and X-ray flares. We therefore address the UV and X-ray flares separately. 

\subsection{Ultraviolet flares}\label{subsec:uvflares}

Following the works of \citet{Loyd2014} and \citet{Loyd2018} we used the \texttt{costools} package provided by STScI in order to flux-calibrate the time-series data. We detected one partial flare from \LTT\ with the COS/G160M grating (Fig.\,\ref{fig:timeseries_LTT1445A}). Unfortunately we did not catch the peak of this flare, only a partial decay at the beginning of an orbit. Without any constraints on the flare peak we cannot constrain the duration or energy of this flare, and we therefore do not provide any further analysis.

In the HST observations of \GJ\ we saw two flares, both with the COS/G130M grating (Fig.\,\ref{fig:timeseries_GJ486}). For the first flare we have data before, during, and after the flare. The second flare is much larger and there is a gap in the data when the flare starts due to a change in \texttt{FP-POS}, however we did capture the flare peak. For neither flare do we observe a complete return to the pre-flare quiescent flux level. For both of \GJ's flares we employed the continuous flare model from \citet{TovarMendoza2022}, which is an upgrade to the piece-wise analytic model for classical flares from \citet{Davenport2014}. The \citet{TovarMendoza2022} flare model is a convolution of a Gaussian function with the sum of two exponential functions. The choice of a Gaussian function is physically motivated by the rapid rise in continuum emission, and the double exponential describes the rapid and then more gradual decay back to quiescence \citep{Hawley1991,Kowalski2013,Hawley2014,Davenport2014,Jackman2018,Jackman2019}. 

\citet{TovarMendoza2022} provide an analytic flare model based on photometric optical data from the Kepler survey. This analytic model can be scaled by an amplitude, a characteristic timescale, and a central time to fit to new flares. We found that the analytic model fits the first \GJ\ flare reasonably well, but did not provide a good fit to the second, larger flare, which appears to have a longer rise timescale. It is possible that because the analytic model was derived using fits to optical flare data, it is not directly applicable to flares at broad UV wavelengths \citep{Berger2023}. \citet{Feinstein2022} also find that the analytic optical model did not match high-energy UV flares detected from the active star AU Mic. We therefore determined a new model fit to each flare following the methods in \citet{TovarMendoza2022}, rather than using the analytic form derived from Kepler photometry.

A full derivation of the flare model can be found in \citet{TovarMendoza2022}, so here we only reproduce the final flare equation, after the convolution of the Gaussian and exponential functions has been applied:

\begin{equation}\label{eqn:flare}
    \text{Flare}(t) = \frac{\sqrt{\pi}AC}{2} 
                        \left[
                            \begin{array}{l}
                            F_1 \left(
                                \begin{array}{l}
                                e^{-D_1t_\text{rel} + \left(\frac{B}{C} + \frac{D_1C}{2}\right)^2} \\
                                \quad \times\ erfc\left(\frac{B-t_\text{rel}}{C}+\frac{D_1C}{2}\right)  
                                \end{array}
                                \right)\\ [5ex]
                           \quad + \quad F_2 \left(
                                \begin{array}{l}
                                e^{-D_2t_\text{rel} + \left(\frac{B}{C} + \frac{D_2C}{2}\right)^2} \\
                                \quad \times\ erfc\left(\frac{B-t_\text{rel}}{C}+\frac{D_2C}{2}\right)  
                                \end{array}
                                \right)\\ [1ex]
                            \end{array}
                       \right]
,\end{equation}

\noindent where

\begin{itemize}
    \item $A$ is the flare amplitude;
    \item $B$ is the central position of the Gaussian function;
    \item $C$ is the Gaussian rise timescale;
    \item $D_1$ is the rapid decay phase timescale;
    \item $D_2$ is the slow decay phase timescale;
    \item $F_2$ is defined as 1 – $F_1$, which describes the relative importance of the exponential decay terms $D_1$ and $D_2$; and
    \item $erfc$ is the complementary error function from \texttt{scipy.special}.
\end{itemize}

\noindent The flare model is a function of relative time $t_\text{rel}$, which is the unitless value defined as

\begin{equation}
    t_\mathrm{rel
} = \frac{t - t_\mathrm{cen}}{t_{1/2}}
,\end{equation}

\noindent where $t$ is time, $t_\text{cen}$ is a central time that is close to the time of the flare peak, though not exactly at the flare peak due to the convolution, and $t_{1/2}$ is the characteristic timescale \citep{Kowalski2013}. This scaling to a relative time is useful when fitting multiple flares because the priors for the flare function variables only need to be defined once, as opposed to being tailored to each flare.

Again following \citet{TovarMendoza2022}, we determined the best-fit flare model for each flare by performing the following steps:
\begin{enumerate}
    \item Subtract off the median out-of-flare flux such that the baseline is at 0.
    \item Fit the flare using the analytic flare model from \citet{TovarMendoza2022} to determine initial flare scaling parameters: amplitude, central time, and characteristic timescale.
    \item Scale the flare by the initial flare scaling parameters and fit for the flare model variables $A$, $B$, $C$, $D_1$, $D_2$, and $F_1$. 
    \item Fix the best fit flare model variables and then again fit for the flare scaling parameters to get the best-fit flare model.
\end{enumerate}

In both flare cases we did not capture enough data post-flare to see the flux return to the nominal pre-flare flux level; we therefore only used the pre-flare flux level to compute the median out-of-flare flux. We used the \texttt{emcee} package built into the \texttt{lmfit} optimizing package to explore the parameter space \citep{Foreman-Mackey2013,Newville2016}. We ran \texttt{emcee} for 50,000 steps with a 5,000 step burn in and determined convergence by checking the integrated auto-correlation time. We report our best-fit values in Table~\ref{tab:flare_GJ486}.

To compare the two flares we detected from \GJ\ with those from other works \citep[e.g.,][]{Loyd2018}, we used the best-fit flare model and the median out-of-flare flux value to compute the absolute energy $E$ (the integrated flux during the flare minus the quiescent flux) and equivalent duration $\delta$ (the flare energy normalized by the quiescent energy) for each flare, using the standard equations for these values \citep{Loyd2018}. Because we did not observe the return to the quiescent flux level after either flare, in both cases we extended the best-fit model past the end of the observations in order to compute the absolute energies and equivalent widths. We also computed flare parameters from the best-fit flare models, such as time of the flare peak, flare amplitude, and flare fwhm (Table~\ref{tab:flare_GJ486}).
\begin{figure}
\resizebox{\hsize}{!}{\includegraphics{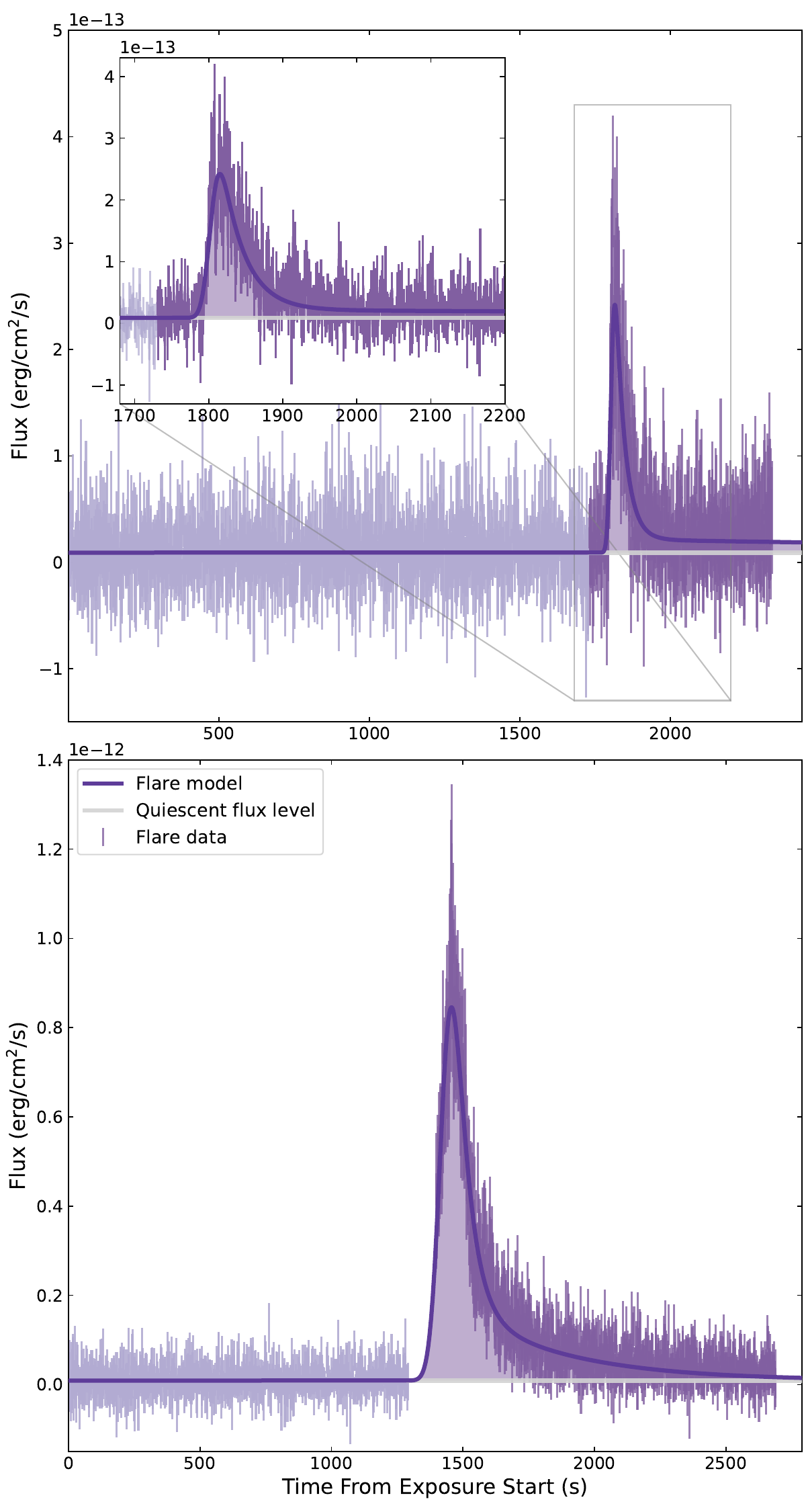}}
\caption{We detected two flares in the time series of the COS/G130M data for GJ 486. Dark purple points highlight the data that we consider to be ``in flare,'' as opposed to the lighter purple ``quiescent'' data. Dark purple lines are the best-fit flare models (Eqn.\,\ref{eqn:flare}, Table~\ref{tab:flare_GJ486}), created from a continuous flare model based on the work of \citet{TovarMendoza2022}. The first flare (Flare 1; \textit{top}) has an absolute energy of 10\tsup{29.5} erg and an equivalent duration of 4357$\pm$96 s, while the second flare (Flare 2; \textit{bottom}) is larger, with an absolute energy of 10\tsup{30.1} erg and an equivalent duration of 19724$\pm$169 s.
} \label{fig:flare_GJ486}%
\end{figure}
\begin{table*}
\centering
\caption{\GJ\ flare properties \label{tab:flare_GJ486}}
\begin{tabular}{l|c|cc}
\hline\hline
 & Unit & Flare 1 & Flare 2 \\
 \hline
Wavelength range               & \AA                    & 1065--1367                        & 1065--1367 \\
Flare start--end\tsup{a}       & s                      & 1730--2161                        & 1300--2690 \\
Flare duration\tsup{a}         & s                      & 431                       & 1390       \\
Median quiescent flux          & erg cm$^{-2}$ s$^{-1}$ & $9.02\times10^{-15}$              & $8.89\times10^{-15}$\\
\hline
& & \multicolumn{2}{c}{Flare model parameters}\\
\hline
$t_\text{cen}$          & s       & $1809.60 \pm 0.80$  & $1438.22 \pm 0.89$          \\         
$t_{1/2}$               & s       & $36.2 \pm 1.8$      & $89.6 \pm 1.7$              \\         
$A$                     & ---     & $1.28 \pm 0.19$         & $1.50 \pm 0.10$        \\
$B$                     & ---     & $-0.182 \pm 0.041$  & $-0.164\pm 0.020$\\
$C$                     & ---     & $0.364 \pm 0.061$   & $0.536 \pm 0.026$ \\
$D_1$                   & ---     & $0.015 \pm 0.017$   & $0.226 \pm 0.014$ \\
$D_2$                   & ---     & $1.16 \pm 0.13$     & $1.8 \pm 0.16$  \\
$F_1$                   & ---     & $0.0327 \pm 0.0058$ & $0.0988 \pm 0.0067$ \\
\hline
& & \multicolumn{2}{c}{Derived flare properties}\\
\hline
Time of peak flare\tsup{a}     & s                      & $1814$                          & $1457$               \\
Flare amplitude                & erg cm$^{-2}$ s$^{-1}$ & $2.34\times10^{-13}$            & $8.36\times10^{-13}$               \\
FWHM                           & s                      & $47$                            & $123$                \\
Absolute energy ($E_\text{COS/G130M}$) & erg            & $3.065\pm0.068 \times 10^{29}$   & $1.368 \pm 0.012 \times 10^{30}$ \\
Equivalent duration ($\delta$) & s                      & $4357\pm96$                     & $19724\pm169$  \\
\hline
\end{tabular}
\tablefoot{A description of the flare model parameters can be found in Sect.\,\ref{subsec:uvflares}. Derived flare properties are computed from the best-fit flare models.\\
\tsup{a} Flare times are relative to the observation start time. We do not observe the start of Flare 2, so time values are approximate.
}
\end{table*}

We additionally inspected how the two flares affect individual transition lines in the FUV observed with the COS/G130M grating. We reiterate that, at the beginning of the analysis, we removed time-series data containing the flares from the rest of the observations used to make the panchromatic spectrum. We used the removed flare data to create FUV spectra and fit the line profiles of observed transition region lines using the same methods as described in Sect.\,\ref{subsec:ultraviolet}. 

\begin{figure*}
\centering
\includegraphics[width=17cm]{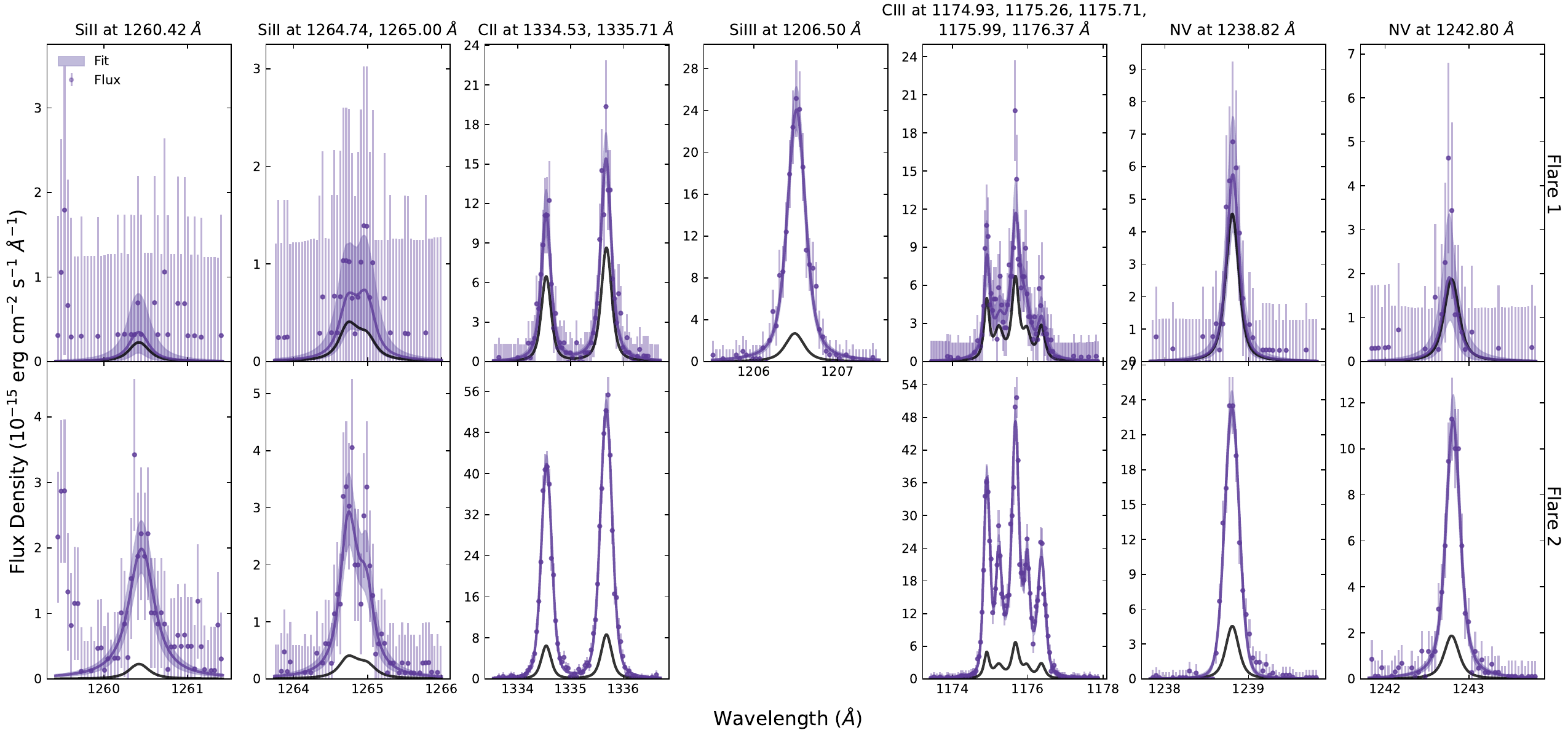}
\caption{Transition region lines detected with COS/G130M for \GJ\ during the two observed flares. The purple line profiles and fits are constructed from data taken during the ``in flare'' phase of each flare. The black line profiles are the quiescent states; they are the same quiescent profiles in the top and bottom rows, but note the change in y-axes. Flare 2 was observed using FP-POS 4, which misses the Si \textsc{iii} line at 1206.5\AA.}
\label{fig:flare_spec_GJ486}
\end{figure*}

We compared the flux density of each line between the quiescent state and each flare state (Fig.\,\ref{fig:flare_spec_GJ486}). The flare data are shown in purple, with purple lines representing the best-fit Voigt line profiles convolved with the COS LSF. These are compared to the best-fit line profiles from the quiescent data (black lines). The transition region lines are ordered from lower to higher formation temperatures. It is not surprising that the more energetic Flare 2 shows a larger increase in individual transition region lines than Flare 1. Note that observations for Flare 2 used FP-POS 4, which does not cover the Si \textsc{iii} line. The greatest increases in line flux are for lines forming at intermediate temperatures of 4.5 $>$ log\tsub{10}(T) $>$ 5.0: C \textsc{ii}, Si \textsc{iii}, and C \textsc{iii} \citep{France2016,Loyd2018}. In Flare 2 we also see emission from the coronal FUV line Fe XXI at 1354\,\AA.

\subsection{X-ray flares}
\label{ssec:xrayFlare}

\LTT\ flared during the 2021 \textit{Chandra} observations and was analyzed in detail in \citet{Brown2022}. We reproduce the flare time series in Fig.\,\ref{fig:Xray_LTT1445A}, along with the quiescent time series taken in 2023 and used in the panchromatic spectrum in this work. The flare flux is about 55$\times$ greater than the average quiescent level just prior to the flare, and about 7$\times$ greater than the quiescent level observed two years later in 2023. The characteristic single-temperature fits to these spectra show significantly hotter plasma present during the flare than in quiescence (see Table \ref{tab:Xray_LTT1445A}).

The LTT 1445A  datasets obtained in ACIS VFAINT mode (Fig.\,\ref{fig:Xray_LTT1445A}) were tested for source variability using the CIAO tool \texttt{glvary}\footnote{glvary CIAO thread: \href{https://cxc.cfa.harvard.edu/ciao/threads/variable/}{cxc.cfa.harvard.edu/ciao/threads/variable/}}. This tool searches for variability using the Gregory-Loredo algorithm \citep{Gregory&Loredo1992}, which tests for nonrandom grouping of the event times across multiple time bins. The \texttt{glvary} tool is regularly used to test for variability in the major \textit{Chandra} source catalogs and provides a variability index (VARINDEX) which ranges from 0 to 10, with values of 5 or above indicating a variable source. The 2021 observation with its large flare has a VARINDEX of 10. The 2023 observation, while less dramatic, is also definitely variable with a VARINDEX of 7. The variability seen in the 2023 dataset seems to be typical for LTT 1445A.

For \GJ\ we observed flares with the \textit{XMM-Newton} EPIC-pn camera, as shown in Fig.\,\ref{fig:Xray_GJ486}. The biggest flare occurred about halfway through the observation, and is clearly visible in the harder energies (bottom panel), with only a smaller, nonsignificant rise in the soft band (middle panel). Our spectral fits in Sect.\,\ref{ssec:xraySpec} indicate a 5.5$\times$ increase in the overall flux during the flare, with the bulk of that rise coming through an increase in the higher of the two temperature components. This ``hardening'' of the X-ray emission is typical of flares at these wavelengths \citep[e.g.,][]{Reale2001,Pye2015}.

Two small peaks in the following $\sim$7\,ks are also possibly associated with flaring, with both rises only seen in the hard band and not the soft. The first is not statistically above the basal level, and was not excluded from the quiescent epoch in our spectral fits in Sect.\,\ref{ssec:xraySpec}. The peak of the second does show a clear significant rise above the quiescent level. The signal was too low to warrant fitting a separate spectrum, but this epoch was not included the quiescent spectrum.

We attempted to search for evidence of the flaring in the simultaneous near-UV data taken with the OM. Looking at the raw image mode count rates shows evidence of a possible rise shortly after the midpoint of the observation. However, the data also show major problems. While much of the background in a single exposure shows $\sim$10 counts per pixel, there are many with $>$10,000 counts spread all across the image. While suggestive of guiding issues and/or the source being too bright, there is still a source visible in the correct position on the image, and the hallmarks of typical OM saturated sources are not present. The problems seen in the OM data lead the later parts of the standard reduction chains to fail. We do not consider this data further.

\subsection{Flare frequency}

Based on the observed flares and Poisson counting statistics we make rough approximates of flare frequencies. Optical activity indicators suggest that both \LTT\ and \GJ\ can be considered inactive stars. From two sectors of TESS data on \GJ, a visual inspection did not reveal any flares. The TESS photometric light curve for \LTT\ is contaminated by an active binary companion $7''$ away; there are many flares in the \LTT\ TESS light curve, but their origin is likely from LTT 1445B and C, and the three components cannot be disentangled to determine the flare origins. 

From the two \GJ\ flares observed in the FUV with the COS/G130M grating (observations that lasted a total of 7610 s), we derive a flare frequency $23\pm5$ flares/day with absolute energy $E\geq10^{29.5}$ erg and equivalent duration $\delta\geq4,300$ s. We compare this estimate directly to work by \citet{Loyd2018}, who used observations with the same COS/G130M grating to perform a statistical analysis of flares in sample of 10 active and inactive M dwarfs. We find that the derived flare frequency for \GJ\ in this work is similar to what we expect for typically active M dwarfs when comparing absolute energies \citep{Loyd2018}. However, \citet{Loyd2018} demonstrated that flare frequencies for inactive and active M dwarfs are statistically indistinguishable when comparing flares in relative units, such as equivalent duration. When comparing the derived flare frequency for \GJ\ to the \citet{Loyd2018} sample we find that the supposedly inactive \GJ\ flares at a rate of more than 10$\times$ that of the stars in the \citet{Loyd2018} sample. For \LTT\ we detected a flare, but because we cannot determine its energy or duration, we cannot derive a UV flare rate.

The relationship between flare energy in the UV and X-ray is not well established, with only limited information available regarding the relative strengths of a single flare event across the electromagnetic spectrum \citep{MacGregor2021}. Simultaneous observations in X-ray and FUV did observe flares on Proxima Centauri and found a coherent increase in the derived DEM corresponding to the X-ray and FUV regimes \citep{Fuhrmeister2022}. While we observed X-ray flares for both \LTT\ and \GJ, these are not the counterparts to the UV flares, which were observed at different times. The X-ray time series data are not flux-calibrated, meaning that we cannot derive X-ray flare frequency rates.

\section{Discussion}\label{sec:discussion}
\subsection{Planetary atmospheres}\label{subsec:planetatmospheres}
A driving motivation for capturing the high-energy flux from exoplanet-hosting M dwarfs and producing self-consistent panchromatic spectra is to gain a holistic understanding of their terrestrial planets' atmospheres. High energy flux from M dwarfs influence both photochemistry and atmospheric escape from the upper atmospheres of orbiting planets \citep{Catling&Kasting2017,Kubyshkina2024}. The photoevaporation theory of atmospheric evolution posits that the terrestrial exoplanets we observe today may have started off with primordial hydrogen-rich atmospheres that were subsequently lost due to hydrodynamic escape \citep{Owen&Wu2013,Lopez2013,Luger2015a,Owen2020b}. Hydrodynamic escape is driven by the absorption of stellar high energy radiation in the upper atmospheres of planets. As hydrogen is driven away, it can drag heavier material along with it, leading to loss of atomic and molecular species such as O, H\tsub{2}O, and CO\tsub{2} \citep{Zahnle&Kasting1986,Pepin1991,Odert2018,Lammer2018}. Photodissociation of H\tsub{2}O can continuously fuel hydrodynamic escape, and in the extreme lead to complete dessication of the planet \citep{Luger2015b}, as perhaps is the case for Venus \citep[e.g., ][]{Gillmann2009}.

Early hydrodynamic escape is rapid, and can last from a few thousand years to a few Myr after formation, depending on a planet's initial water content and atmospheric composition, and the stellar high energy flux. Hyrodynamic escape can result in the complete loss of atmospheric atomic hydrogen, as well as losses of a few to a few thousands of bars of heavier species, again depending on the initial conditions. For example, in a study focused on the terrestrial exoplanet TRAPPIST-1c orbiting a late-type M dwarf, \citet{Teixeira2024} impose initial water contents of 0.01 to 100 Earth oceans on TRAPPIST-1c, and find that hydrodynamic escape can last for 1000 years to 10 Myr resulting in a loss of CO\tsub{2} ranging from 0.1 to 1000 bars.

The measured radii and masses of the terrestrial planets \LTTb, \LTTc, and \GJb\ result in high bulk densities that are inconsistent with H/He-rich atmospheres \citep{Trifonov2021,Winters2022}, and, based on hydrodynamic escape models, we do not expect these worlds to currently retain primordial H/He-dominated atmospheres accreted from the protoplanetary nebula \citep{Kubyshkina2021}. It is possible, however, that these worlds have secondary atmospheres consisting of heavier molecular material than hydrogen that is not so easily lost. Secondary atmospheres can arise from material that was sequestered in the mantle during formation and outgassed later on timescales of a few Gyr \citep[e.g.,][]{Marty&Dauphas2003}, or else delivered by comets or asteroids \citep{Raymond2007,Ciesla2015,OBrien2018}. Secondary atmospheres can still be sculpted and removed by stellar wind, which is particularly relevant for terrestrial exoplanets on short orbital periods around M dwarf stars \citep[e.g.,][]{Garraffo2017}. It is also possible that \LTTb, \LTTc, and \GJb\ are airless rocks \citep[e.g.,][]{Kreidberg2019,Crossfield2022,Zieba2023,Zhang2024}.

To connect the measured X-ray and EUV radiation from the M dwarfs \LTT\ and \GJ\ to the atmospheric mass loss of their terrestrial exoplanets, we employ a toy model following the steps of \citep{Teixeira2024}. We skip the rapid hydrodynamic escape phase since the mass loss is heavily dependent on initial water content, for which we do not have accurate estimates, and it is clear from radius-mass measurements that \LTTb, \LTTc, and \GJb\ do not have hydrogen-dominated atmospheres. We instead focus on whether or not pure CO\tsub{2} secondary atmospheres on these worlds can survive to the present day in the presence of stellar wind, though atmospheres can also be sculpted and stripped over long timescales via thermal processes such as Jeans escape \citep{VanLooveren2024}. Our toy model does not include rates of mantle outgassing or the possibility of magnetic fields on the planets, which can shield atmospheres from stellar ions.

We use the X-ray luminosity--age relationship developed by \citet{Engle2024} for M2.5--6.5 dwarfs to get an estimate of the X-ray luminosity over time for the two targets, up until their current estimated ages (details on stellar ages to follow in Sect.\,\ref{subsec:starage}). We then calculate the stellar mass loss rates following \citet{Teixeira2024}, who derive their stellar mass-loss rate relationship from \citet{Wood2021}:

\begin{equation}
    \dot{M_\text{s}} = \frac{\mathcal{A}}{10^{3.42}}F_\text{X,surf}^{0.77}
,\end{equation}

\noindent where $\mathcal{A}$ is the stellar surface area in solar units, and $F_\text{X,surf}$ is the X-ray surface flux of the star. The result is the stellar mass loss rate in units of solar mass loss rate, which we take as $1.26\times10^{12}\,\text{g/s}$.

From the stellar mass loss rate we calculate the planetary atmospheric mass loss rate due to stellar wind stripping as

\begin{equation}
    \dot{M}_\text{atm} = 0.025 \left(\frac{R_\text{p}}{a}\right)^2 \dot{M_\text{s}}
,\end{equation}

\noindent where $R_\text{p}$ is the planetary radius and $a$ is the planetary semi-major axis \citep{Dong2018,Teixeira2024}. The factor of 0.025 is borrowed from the TRAPPIST-1 system because we do not have MHD models of \LTT\ or \GJ; however they are also M dwarfs and their planetary orbital distances are similar to those of the TRAPPIST-1 planets \citep{Agol2021}.

When executing the toy models we assume pure \C\ atmospheres for \LTTb, \LTTc, and \GJb, since \C\ is a common atmospheric molecule with advantages for atmospheric retention: it has a high mean molecular weight and a rapid cooling timescale that makes it resilient to thermal escape \citep{Gordiets1985,Johnstone2021,VanLooveren2024}. For each planet we start with a pure CO\tsub{2} atmosphere equal to 5--50\% of Earth's total estimated CO\tsub{2} budget of 10\tsup{22} mol \citep{Sleep&Zahnle2001,Foley&Smye2018}. Though we do not take into account outgassing rates, \cite{Teixeira2024} find that most of the outgassing takes place in the first Gyr. For each time step, we compute the planetary atmospheric mass loss rate from the stellar mass loss rate, and subtract the result from the remaining CO\tsub{2} in the atmosphere.

\begin{figure}
\resizebox{\hsize}{!}{\includegraphics{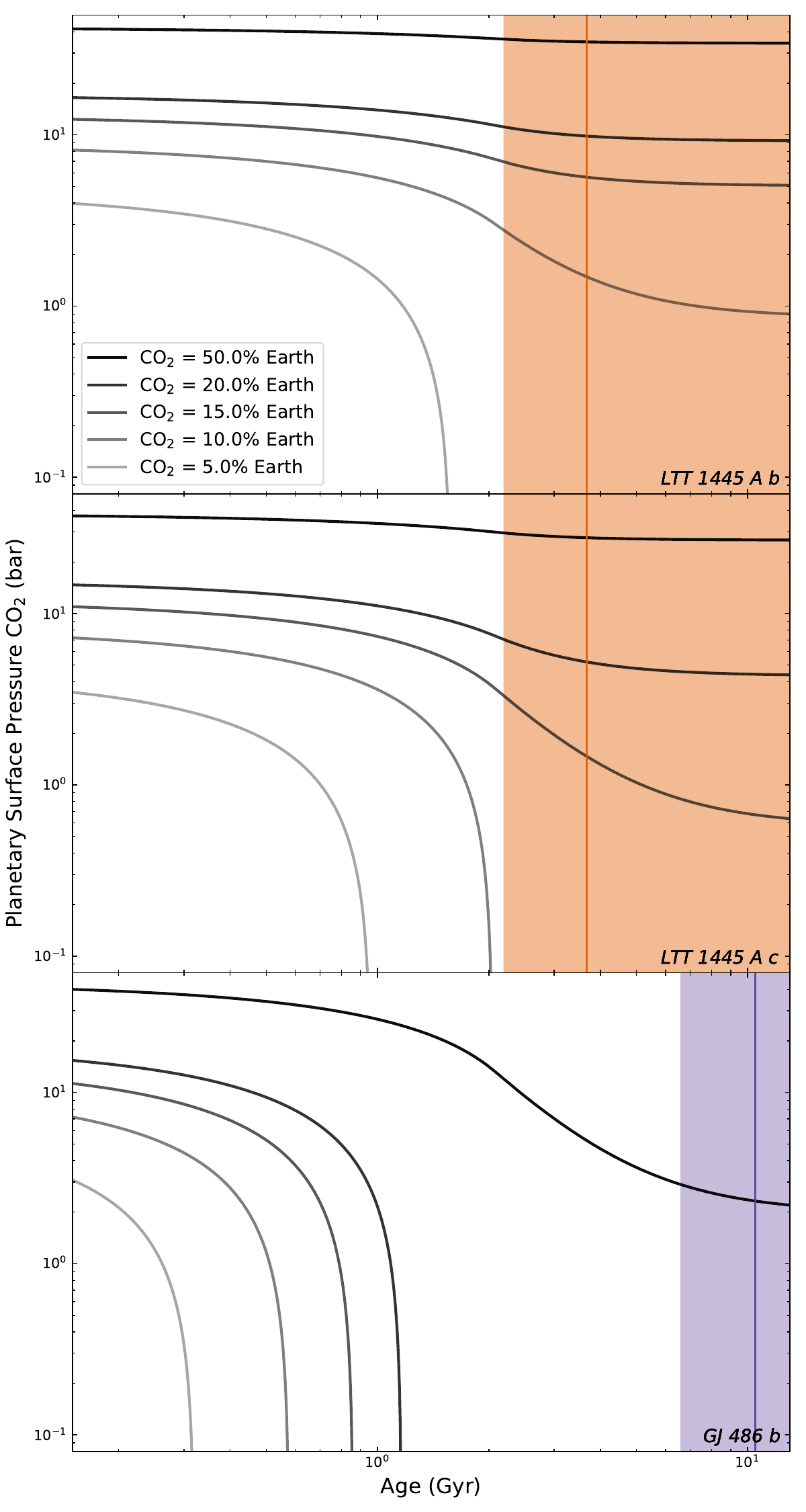}}
\caption{Results of the toy model describing \C\ mass loss from each terrestrial planet, assuming initial atmospheres of pure \C\ 5--50\% that of Earth's total estimated \C\ content of $10^{22}$ mol (Sect.\,\ref{subsec:planetatmospheres}). The vertical orange and purple lines, along with the shaded regions, are the estimated stellar ages with uncertainties (discussed in Sect.\,\ref{subsec:starage}). Whether or not each planet maintains a \C\ atmosphere with surface pressure of $\gtrapprox1$ bar depends on the initial \C\ content, the planetary orbital period, and the stellar age and activity history. The model does not take into account \C\ outgassing rates---we assume all of the \C\ outgasses at the start of the model time steps---or magnetic fields, which may be able to shield secondary atmospheres from loss.} \label{fig:planetmassloss}%
\end{figure}

We present the results of our toy model in Fig.\,\ref{fig:planetmassloss}. Whether or not the planets \LTTb, \LTTc, and \GJb, can retain CO\tsub{2}-rich atmospheres is dependent on their initial atmospheric CO\tsub{2} content, their orbital distances, and the age of their host stars. So long as \LTTb\ has an initial CO\tsub{2} budget of at least 10\% that of Earth's, it should be able to maintain a CO\tsub{2} atmosphere of at least 1 bar of surface pressure. \LTTc, which orbits closer to the host star, needs at least 15\% of Earth's CO\tsub{2} budget to maintain an atmosphere of almost 1 bar of surface pressure. Finally, \GJb\ orbits the closest to its host star, making its atmosphere the most susceptible to loss; \GJb\ needs an initial \C\ content of at least 50\% of Earth's current \C\ content to maintain a \C-rich atmosphere of at least 1 bar of surface pressure.

Without knowing the initial compositions of \LTTb, \LTTc, and \GJb\ or the activity history of their host stars, we cannot know \textit{a priori} whether or not they have atmospheres today. Determining whether or not these planets have atmospheres can place upper limits on their initial water contents \citep[e.g.][]{Kreidberg2019}. One scaling relation called the ``cosmic shoreline'' compares the (estimated) cumulative XUV radiation received by a planet and its escape velocity under assumptions of energy-limited escape \citep{Zahnle2017}. All three planets considered in this work lie close to the $I_\text{XUV} \propto v_\text{esc}^4$ relationship, implying that all three planets have undergone hydrogen-stripping but are not necessarily bare rocks. Observational programs looking at \LTTb\ and \GJb\ in secondary eclipse with JWST/MIRI LRS spectroscopy (GO Program 1743, PI Mansfield; and 2807 PI Berta-Thompson) and in transmission with JWST/NIRSpec BOTS (GO Program 2512, PI Batalha) may soon provide the answer for these two worlds.

\subsection{Stellar ages}\label{subsec:starage}
It is notoriously difficult to determine the ages of M dwarfs. They spend hundreds of Gyr on the main sequence \citep{Choi2016}, and once they are stably fusing hydrogen in their cores, their measurable properties like luminosity, effective temperature, and radius remain constant on cosmological timescales; in other words, longer than the age of the universe. This means that stellar evolutionary models are poor predictors of M dwarf age, even with well-constrained stellar properties in the \textit{Gaia} era \citep{GaiaMission2016,GaiaDR32023,Eastman2023}.

One promising avenue is to apply the age--rotation or age--activity relationships that exist for main sequence stars, and set constraints specifically for M dwarfs \citep{Wright2016,Wright2018}. The fact that older main sequence stars have longer rotation periods and decreased activity relative to younger main sequence stars is well-documented \citep[e.g., ][]{Schatzman1962,Skumanich1972,Barnes2003,Newton2016a,Pineda2021}. Using a series of dating methods to pinpoint the ages of a precious sample of M dwarfs, \citet{Engle2023} establish age--rotation relationships for early- to mid-M dwarfs, which are then expanded to age--activity relationships \citep{Engle2024}. Here we apply some of these relationships to \LTT\ and \GJ.

Both \LTT\ and \GJ\ have long rotation periods: $85\pm22$ and $130.1\pm1.6$ days, respectively \citep{Trifonov2021,Winters2022}; and low activity: only weak measurements of H$\alpha$ and Ca \textsc{ii} H \& K \citep{Astudillo-Defru2017,Hojjaptpanah2019,Winters2019}. These long rotation periods and weak activity indicators broadly put \LTT\ and \GJ\ in the category of ``old, inactive'' M dwarfs, but here we attempt to derive more quantitative ages. Using the age--rotation relationships for M2.5--6.5 dwarfs with rotation period greater than 24 days, we estimate that the ages of \LTT\ and \GJ\ are $5\pm\text{Univ}$ and $7.8\pm4.8$ Gyr, respectively \citet{Engle2023}, where we use the limit of ``Univ'' to mean that the uncertainties extend to the age of the universe. We also note that additional work on \GJ\ finds a significantly shorter stellar rotation period of $49.9\pm5.5$ days \citep{Caballero2022}, which translates to a stellar age of $3.7^{+5.1}_{-\text{Univ}}$ Gyr \citep{Engle2023}.

The unconstrained age of \LTT\ is due to the difficulty in measuring its rotation period; the value of $P_\text{rot}=85\pm22$ days is derived from the mass-rotation period relation for inactive M dwarfs \citep{Newton2017}. \LTT\ is the primary member of a hierarchical stellar triple, with a binary pair roughly 7$''$ away. The binary stars are more active and have rapid rotation periods, making it difficult to distinguish the much longer rotation period of \LTT\ \citep{Winters2022}. Based on an analysis of galactic kinematics and flare rates of a volume complete sample of mid-to-late M dwarfs, \citep{Medina2022b} determine broad age bins based on stellar rotation period: mid-to-late M dwarfs with $10<P_\text{rot}<90$ days are $5.6\pm2.7$ Gyr, and those with $P_\text{rot}>90$ days are $12.9\pm3.5$ Gyr. These age-bin estimates agree with those found using the age--rotation relation from \citet{Engle2023}. 

We are better able to constrain the ages of \LTT\ and \GJ\ using their high-energy spectra and age--activity relationships \citep{Engle2024}. We opt to use the log($L_\text{X-UV (5--1700\AA)}/L_\text{bol}$) activity--age relation because this is the best defined in terms of wavelength range. We integrate the panchromatic spectra from 5--1700\,\AA\ and find that \LTT\ has an age of $3.7^{+\text{Univ}}_{-1.5}$ Gyr and \GJ\ has an age of $10.5^{+\text{Univ}}_{-3.9}$ Gyr. The upper and lower bounds of the stellar age are conservative: we take the integrated X-UV (5--1700\AA) flux plus and minus 1$\sigma$ errors on the panchromatic spectrum and compare these against the 1$\sigma$ errors in the activity--age relationship from \citet{Engle2024}. We take the outer-most ages that take into account the uncertainty in the integrated spectrum and the activity--age relation. We place lower limits on the ages of \LTT\ and \GJ\ at 2.2 and 6.6 Gyr, respectively. 

Conservatively, we cannot put an upper bound on the age of \LTT\ and \GJ; however the $1\sigma$ lower age bounds are still useful. For example, based on their flare-rate analysis as a function stellar rotation period, \citet{Medina2022b} determine a transition age for mid-to-late M dwarfs from a saturated to unsaturated flaring state at $2.4\pm0.3$ Gyr, though we note that this result is based on optical flares in the TESS data. If this transition age is true for UV flares, then \GJ\ is in a state of decreased activity and relatively low flaring, while \LTT\ may be in the process of transitioning between the saturated and decreasing activity states. The UV flare rate we deduce for \GJ\ is significantly higher than what we expect for an inactive star, suggesting that high-energy flare rates may remain high as optical flare rates decrease.

\subsection{Evidence of prolonged elevated activity} \label{subsec:stellar_activity}

We do not have access to long-term high-cadence monitoring of a large sample of inactive M dwarfs at high-energy wavelengths. Rather, what we present here are high-cadence observations in spurts from HST. Based on these noncontiguous observations, we posit that supposedly ``inactive'' M dwarfs like \LTT\ and \GJ\ not only produce more high-energy flares per day than in the optical, but also that those flares produce prolonged periods of heightened flux output, perhaps lasting as long as several days, and may be the result of observing during a high-activity period. Here we present evidence for this conjecture. 

We look first at \GJ\ because for this star we observe two flares and are able to model them and derive their absolute energies and equivalent durations. It is well-documented that stars that are considered inactive at optical wavelengths still flare in the FUV \citep{Loyd2018,Froning2019,Loyd2020,Diamond-Lowe2021,Jackman2024}. The star GJ 674 is considered inactive with a rotation period of 30 days, no H$\alpha$ in emission, and an approximate age of 0.1 to a few Gyr \citep{Bonfils2007,Newton2018}, and yet multiple flares, the largest having an absolute energy of $E=10^{30.75}$ erg and an equivalent duration of $>30,000$ s, were observed up by HST COS/G130M \citep{Froning2019}. By comparison, \GJ\ has an even longer rotation period of 130 days and a lower age limit of 6.6 Gyr, and two flares with absolute energies $E=10^{29.5}$ and $E=10^{30.1}$ erg and equivalent durations of $4357\pm96$ and $19724\pm169$ s were observed with the same grating within three hours. Old and quiet M dwarfs like \GJ\ can still produce high energy flares, but what is surprising is that we derive a flare frequency roughly 10$\times$ that of the broader M dwarf population \citep{Loyd2018}, with the caveat that we are very much in the regime of small sample statistics based on the limited time-sampling of our observations. We find it more plausible that we caught \GJ\ in a period of heightened flare activity than this being the true flare frequency for this star.

We additionally posit that this period of heightened activity in \GJ\ lasts at least a week. We observed the two flares from \GJ\ in the same HST visit, only three hours apart (Fig.\,\ref{fig:timeseries_GJ486}), perhaps suggesting that both flare events originate from the same activated region of the stellar surface. We showed in Sect.\,\ref{subsec:lya} and Fig.\,\ref{fig:Lya_GJ486} that even after removing the flare data from the COS/G130M observations, the UV-UV line correlations with \Lya\ suggest that lines measured from the ``quiescent'' time series were in fact excited. The next set of UV observations of \GJ\, taken six days later with COS/G160M and COS/G230L, also appear systematically higher that what we estimate for the \Lya\ flux using the reconstruction method \citep{Youngblood2016}. The STIS/G140M observations that provide the data for the \Lya\ reconstruction were taken almost exactly one year apart---three months before the COS/G130M observations were we observe the flares and eight months after the COS/G160M and COS/G130M observations---yet they agree incredibly well. Our final line of evidence for prolonged activity comes from the \LTT\ data, where we observed the tail end of a flare with COS/G160M. We do not know how big this flare was, but two days later we observed \LTT\ with STIS/G140M, and this observation gives 1.5$\times$ higher flux than STIS/G140M observations taken six weeks before (Fig.\,\ref{fig:timeseries_LTT1445A}). It is of course also possible that these older M dwarfs might exhibit sporadic rather than periodic activity. Indeed, previous searches did not find strong evidence for persistent flare periodicity in TESS data, though short-term periodic cycles may exist \citep{Howard2021}.

If there are indeed times of heightened stellar activity, whether periodic or not, in older quiet M stars, this affects our ability to determine their ages based on age--activity relations. X-ray and UV measurements of stars generally come from a handful of observations. Wherever a star happens to be in its activity cycle when we observe it will affect the resulting estimated age when using age--activity relations. A good example of this are the three X-ray states---``quiescent,'' ``flare,'' and ``elevated''---of \LTT\ reported by \citet{Brown2022}, and the resulting placement of the star on the X-ray--age relationship based on each state (Fig.\,4 of that paper). Poor time-sampling of high-energy stellar observations are likely contributing to the scatter in the activity--age relationships at both X-ray and UV wavelengths.

Ultimately, we have exhausted the detective work we can do with our noncontiguous and nonoverlapping observations. Additional HST campaigns to capture more flares are risky---even with these observations, \LTT\ and \GJ\ do not flare as reliably as a younger, truly active stars like AU Mic \citep{Feinstein2022}. What is needed is a space mission with a high-energy monitoring campaign of a large catalog of M dwarfs. Consistent time-sampling of M dwarfs at high energy would allow us to determine the periodic nature of activity on older, quiet stars, as well as tighten constraints on age--activity--rotation relationships for M dwarf stars, which would in turn provide the most accurate stellar ages we are likely to get for these long-lived low-mass stars. 

There are proposed missions to monitor large samples of stars, including M dwarfs, in the X-ray \citep[Advanced X-ray Imaging Satellite, or AXIS;][]{Corrales2023} and the EUV \citep[Extreme-ultraviolet Stellar Characterization for Atmospheric Physics and Evolution, or ESCAPE;][]{France2022}. The ESCAPE mission in particular has the goal of EUV monitoring for a sample of M dwarfs with direct application to understanding terrestrial exoplanet atmospheres and the impacts of stellar variability on those atmospheres. The impact of M dwarf activity on observations of terrestrial exoplanets is becoming increasingly important in the JWST era due to stellar contamination of transmission spectra \citep[e.g.,][]{Lustig-Yaeger2023,Moran2023,May2023,Lim2023}. It will take a statistical monitoring survey of M dwarfs at high-energy to push our understanding M dwarf activity periodicity, ages, and impact on terrestrial exoplanet atmospheres to the point where we can eventually answer questions of exoplanet habitability.

\section{Conclusions}\label{sec:conclusions}

In this work, we combine data from HST/COS and STIS (GO Programs 16722 and 16701), \textit{XMM-Newton} (ObsID 23377), and \textit{Chandra} (ObsIDs 23377, 27882, 26210, 27799, 27942), as well as interpolated BT-Settl (CIFIST) models to produce self-consistent panchromatic spectra of \LTT\ and \GJ\ from 1\AA--20$\mu$m. To fill in an observational gap at EUV wavelengths, we use the DEM method to estimate EUV flux \citep{Duvvuri2021}, and we use the wings of the \Lya\ line to reconstruct the full profile \citep{Youngblood2016}. 

Both \LTT\ and \GJ\ are considered inactive based on the long rotation periods and optical activity indicators, but we detect flares in both stars at UV and X-ray wavelengths, suggesting a flare frequency rate that is higher than predicted for inactive stars. The flare frequency distribution we derive from the two UV flares detected from \GJ\ based on their equivalent durations is approximately ten times larger than the larger population of active and inactive M dwarfs \citep{Loyd2018}, despite the fact that this star is optically quiet \citep{Trifonov2021}. The high-energy activity we see in both M dwarfs, despite a lack of activity from optical indicators, is in line with the conclusions of previous works showing that optical observations fail to predict high-energy activity \citep{Loyd2018,Jackman2024}. We consider that this high-energy activity is variable, and may not be captured in a single measurement. A lack of high-energy monitoring of M dwarfs is therefore preventing us from obtaining a comprehensive view of M dwarf activity. We do not know whether the high-energy activity we observe is periodic or sporadic. Only taking snapshots of stars at high energy will also contaminate age--activity relations, which are some of the best ways to constrain M dwarf ages \citep{Wright2016,Engle2024}. 

Variable high-energy activity will also affect our interpretation of planetary atmospheres. \LTTb, \LTTc, and \GJb\ have lost any primordial hydrogen-dominated atmospheres they may have had, but higher mean-molecular-weight secondary atmospheres are still possible. Efforts to determine the atmospheric status of a sample of terrestrial exoplanets orbiting M dwarfs is underway (e.g., the Hot Rocks Survey, JWST GO Program 3730; PI H.\ Diamond-Lowe, Co-PI J.\ M.\ Mendon\c{c}a), with a top recommendation by the Working Group on Strategic Exoplanet Initiatives with HST and JWST to ``understand the prevalence and diversity of atmospheres on rocky-M dwarf worlds'' \citep{Redfield2024}. As noted by the working group report, detecting and characterizing terrestrial exoplanet atmospheres must go side by side with characterizing their M dwarf hosts. Measuring and monitoring high-energy flux from M dwarfs is the next step toward determining the ages, mass-loss rates, and activity cycles of these key exoplanet hosts.

\begin{acknowledgements}
This paper includes data gathered with the Cosmic Origins Spectrograph and the Space Telescope Imaging Spectrograph on board the \textit{Hubble Space Telescope}. This research has used data obtained by the \textit{Chandra X-ray Observatory} and software provided by the \textit{Chandra} X-ray Center (CXC) in the CIAO application package. We make use of the CHIANTI database and open-source python code. CHIANTI is a collaborative project involving George Mason University, University of Michigan (USA), University of Cambridge (UK), and NASA Goddard Space Flight Center (USA). This publication makes use of VOSA, developed under the Spanish Virtual Observatory (\href{https://svo.cab.inta-csic.es}{https://svo.cab.inta-csic.es}) project funded by MCIN/AEI/10.13039/501100011033/ through grant PID2020-112949GB-I00. VOSA has been partially updated by using funding from the European Union's Horizon 2020 Research and Innovation Programme, under Grant Agreement no.\ 776403 (EXOPLANETS-A). This work has made use of data from the European Space Agency (ESA) mission {\it Gaia} \url{https://www.cosmos.esa.int/gaia}), processed by the {\it Gaia} Data Processing and Analysis Consortium (DPAC,
\url{https://www.cosmos.esa.int/web/gaia/dpac/consortium}). Funding for the DPAC has been provided by national institutions, in particular the institutions participating in the {\it Gaia} Multilateral Agreement. This job has made use of the Python package GaiaXPy, developed and maintained by members of the \textit{Gaia} Data Processing and Analysis Consortium (DPAC), and in particular, Coordination Unit 5 (CU5), and the Data Processing Centre located at the Institute of Astronomy, Cambridge, UK (DPCI).

We thank Thea Kozakis for participation in the proposal for GO Program 16722, and Daria Kubyshkina for a helpful discussion of hydrodynamic mass loss. We thank the anonymous referee for their comments, which led to improvements in this paper. H.D.-L.\ acknowledges support from the Carlsberg Foundation, grant CF22-1254. G.W.K.\ and L.C.\ acknowledge support through NASA grant HST-GO-16722.002-A. A.Y. acknowledges support through NASA grant HST-GO-16701.001-A administered by Space Telescope Science Institute. A.B.\ acknowledges \textit{Chandra} grants GO1-22005X and GO2-23002X to the University of Colorado. Y.M.\ acknowledges funding from the European Research Council (ERC) under the European Union's Horizon 2020 research and innovation programme (grant agreement no. 101088557, N-GINE). C.S.\ acknowledges grant DLR 50 OR 2205. \\

Facilities used in this work are: HST (STIS, COS); CXO (ACIS, HRI); XMM-Newton (EPIC).\\

Software used in this work: astropy \citep{AstropyCollaboration2018}; CIAO \citep{CIAO06}; \texttt{dynesty} \citep{Speagle2020}; \texttt{emcee} \citep{Foreman-Mackey2013}; \texttt{lmfit} \citep{Newville2016}; XSPEC \citep{Arnaud1996} \\

\end{acknowledgements}

\bibliographystyle{aa} 
\bibliography{MasterBibliography} 

\end{document}